\documentclass[10pt,final,twocolumn]{IEEEtran}
\ifCLASSINFOpdf
\else
\fi
\usepackage{cite}
\usepackage{amssymb}
\usepackage{amsmath}
\usepackage{graphicx}
\newtheorem{assumption}{Assumption}
\newtheorem{lemma}{Lemma}
\newtheorem{theorem}{Theorem}

\newtheorem{remark}{Remark}

\newtheorem{corollary}{Corollary}

\usepackage{epstopdf}
\usepackage{color}
\usepackage{cases}

\usepackage{subfigure}

\usepackage{bm}

\usepackage{algorithm}
\usepackage{algorithmicx}
\usepackage[colorlinks=true,citecolor=red]{hyperref}

\begin{document}
\title{Online Game with Time-Varying Coupled Inequality Constraints}

\author{
Min Meng, Xiuxian Li, Yiguang Hong, Jie Chen, and Long Wang% <-this % stops a space
%\thanks{This work was partially supported by National Key Research and Development Program of China under Grant 2022ZD0119702, Shanghai Pujiang Program under Grant 21PJ1413100, the National Natural Science Foundation of China under Grant 62103305, 61733018 and 62088101, and Shanghai Municipal Science and Technology Major Project under Grant 2021SHZDZX0100.
	%and the Basic Science Centre Program by the National Natural Science Foundation of China under Grant 62088101.(\emph{Corresponding author: Xiuxian Li.})}
\thanks{M. Meng, X. Li, Y. Hong, and J. Chen are with the Department of Control Science and Engineering, College of Electronics and Information Engineering, and Shanghai Research Institute for Intelligent Autonomous Systems, Tongji University, Shanghai, China (email:  mengmin@tongji.edu.cn); xli@tongji.edu.cn; yghong@tongji.edu.cn; chenjie206@tongji.edu.cn).

%M. Meng and X. Li are also with the Institute for Advanced Study and Shanghai Institute of Intelligent Science and Technology, Tongji University, Shanghai, China.

L. Wang is with the Center for Systems and Control, College of Engineering, Peking University, Beijing, China (email: longwang@pku.edu.cn).
}
% <-this % stops a space
%\thanks{Corresponding author: }
% <-this % stops a space
}

\maketitle

\begin{abstract}
In this paper, online game is studied, where at each time, a group of players aim at selfishly minimizing their own time-varying cost function simultaneously subject to time-varying coupled constraints and local feasible set constraints. Only local cost functions and local constraints are available to individual players, who can share limited information with their neighbors through a fixed and connected graph. In addition, players have no prior knowledge of future cost functions and future local constraint functions. In this setting, a novel decentralized online learning algorithm is devised based on mirror descent and a primal-dual strategy. The proposed algorithm can achieve sublinearly bounded regrets and constraint violation by appropriately choosing decaying stepsizes. Furthermore, it is shown that the generated sequence of play by the designed algorithm can converge to the variational GNE of a strongly monotone game, to which the online game converges.
Additionally, a payoff-based case, i.e., in a bandit feedback setting, is also considered and a new payoff-based learning policy is devised to generate sublinear regrets and constraint violation.
Finally, the obtained theoretical results are corroborated by numerical simulations.
\end{abstract}

\begin{IEEEkeywords}
Decentralized online learning, generalized Nash equilibrium, online game, bandit feedback.
\end{IEEEkeywords}

% \fi
\IEEEpeerreviewmaketitle

\section{Introduction}
Game theory has received growing attention recently owing to its wide applications in social networks \cite{ghaderi2014opinion}, sensor networks \cite{stankovic2012distributed}, smart grid \cite{saad2012game}, and so on. In noncooperative games, the concept of \emph{Nash equilibrium} (NE) plays a pivotal role by providing a rigorous mathematical characterization of the stable and desirable states, from which rational players have no incentive to deviate \cite{basar1999dynamic}.

A challenge is to design decentralized algorithms for seeking NE in noncooperative games based on limited information available to each player. Generally, it is assumed that a coordinator exists to broadcast the data to players \cite{facchinei2010generalized,yu2017distributed,shamma2005dynamic}, that is, bidirectional communication with all the agents is required, which results in high communication loads and is impractical for many applications. Therefore, decentralized algorithms on computing NEs in noncooperative games without full action information, that is, in a partial decision information setting, have been getting more and more attention in recent years. To deal with such kind of scenarios, numerous results on the NE seeking problems have sprung up both in continuous-time \cite{de2019distributed,gadjov2019passivity} and in discrete-time \cite{koshal2016distributed,salehisadaghiani2019distributed,tatarenko2019geometric}, where the algorithms were designed based on gradient descent and consensus schemes.
%The algorithms in \cite{salehisadaghiani2019distributed} using fixed stepsize schemes may have a faster convergence rate than those in \cite{koshal2016distributed} equipped with vanishing stepsizes. As an NE of a convex game can be equivalently expressed as a zero point of a monotone operator, the authors of \cite{tatarenko2018geometric,pavel2020distributed} proposed decentralized algorithms for solving the NE seeking problem by the operator theoretic theory.

All the references mentioned above assumed that all players are rational, which, however, may not be practical particularly in uncertain, unpredictable or adversarial environments. Usually, the players only know they are involved in a repeated decision progress where a reward is returned after an action is made, while they have no knowledge on how the game generates the reward and they do not even know they are playing a game. In this regard, a player must adapt other players' strategies in a dynamic manner by minimizing its (static or dynamic) {\em regret}, i.e., the cumulative payoff difference between a player's actual policy and the best (fixed or time-varying) action in hindsight, which is commonly regarded as a concise and meaningful benchmark for quantifying the ability of an online algorithm for decision-making in the presence of uncertainties and unpredictabilities \cite{cesa2006prediction}. Along this line, mirror descent-based learning algorithms can achieve an order-optimal static regret bound, i.e., $\mathcal{O}(\sqrt{T})$, when cost functions of players are convex \cite{abernethyoptimal}.

It should be noted that the case where the game participated in by players remains unchanged throughout the learning process is most investigated. However, dynamic environments appear frequently in a multitude of practical applications, such as allocating radio resources and online auction, and then the games or the cost functions of players in games often vary with time due to exogenous variations. For example, in a repeated online auction where resources are auctioned off for several bidders, if the gain that each bidder gets for a unit of resource changes over time, then the bidders' utilities are also time-varying. For this kind of online game, the authors in \cite{duvocelle2022multiagent} studied the equilibrium tracking problem and analyzed convergence properties of mirror descent-based policies. Furthermore, a no-regret algorithm based on projection strategies and primal-dual methods was devised in \cite{lu2020online} for online game with time-invariant nonlinear constraints.

It is well known that no-regret learning may not imply convergence to NEs even for the mixed-strategy learning in finite games. In fact, the empirical frequency of play generated by no-regret learning in finite games converges to the game's set of coarse correlated equilibria, which may contain inadmissible strategies, such as the ones that assign positive weight only to dominated strategies\cite{viossat2013no}. The players' learning policy may cycle even in two-player zero-sum games \cite{mertikopoulos2018cycles} and may also lead to an unpredictable and chaotic scenario. It has been proven in \cite{heliou2017learning} that the sequence of play under no-regret learning for potential games converges, including the bandit feedback case. Therefore, it is not explicit to establish the relationship between the no-regret learning and convergence to NEs if no further assumptions or specialized techniques especially for online game with coupled constraints.

%All the references mentioned above considered offline games, where both the cost functions and constraints are time-invariant. However, dynamic environments always exist in a multitude of practical applications, such as allocating radio resources and online auction. In this scenario, cost functions and constraints in a game are time-varying and their values and gradients can be accessible only after decisions are made at the current time. These motivate researchers to find distributed online or learning algorithms for seeking NEs or generalized NEs (GNEs). Along this line, the authors of \cite{lu2020online} applied a primal-dual strategy and consensus algorithms to devise a no-regret online algorithm for seeking GNE of a time-varying game, where cost functions are time-varying while nonlinear constraints are invariant.

On the other hand, this issue becomes considerably intricate when at each round only the reward is revealed for an action profile, instead of the gradients of cost functions or the entire knowledge of cost functions. Bandit algorithms are suitable to solve this kind of problems, including optimization \cite{cao2019online,yi2020distributed_b} and NE seeking \cite{bravo2018bandit,mertikopoulos2019learning}. As to repeated games, it is usually assumed that actions are synchronized across players and rewards are assumed to arrive instantaneously, and then the players need to infer gradient information from a single point observation at each round.

In this paper, we further investigate online game, where players selfishly and simultaneously minimize their own time-varying cost functions subject to local feasible set constraints and time-varying coupled nonlinear constraints. In the setup of full information feedback, i.e., both function values and gradient information are available, a decentralized online algorithm for the studied online game based on mirror descent and primal-dual schemes is devised, and is rigorously proved to ensure that the regrets and constraint violation grow sublinearly by appropriately choosing decaying stepsizes. Further analysis on the convergence of the generated sequence of play by the designed algorithm is provided.
In addition, in view of the difficulty to compute the gradients of cost functions and constraint functions, a payoff-based online learning algorithm relying on one-point stochastic approximation is proposed, and the corresponding analysis is provided to show that the algorithm can achieve sublinear expected regrets and constraint violation.
The main contributions of this paper are summarized as follows.
\begin{itemize}
\item[1)] A decentralized no-regret algorithm is proposed for online game with time-varying shared nonlinear constraints, while the authors in \cite{facchinei2010generalized}--\!\!\cite{tatarenko2019geometric} only studied static games. Existing no-regret algorithms for online game consider either no constraints \cite{duvocelle2022multiagent} or time-invariant constraints \cite{lu2020online}.
\item[2)] A challenging issue is to analyze the convergence of the generated sequence of strategies under no-regret learning even for static games. In this paper, strict analysis is provided to show that if the online game converges to a strongly monotone game, then the generated sequence of play by the designed algorithm can converge to the variational GNE of the game limit. Moreover, the convergence rate of the averaged strategy sequence is derived.
%Moreover, if the time-varying game evolves with time but does not converge to a static game, then the generated sequence of play can averagely converge to the evolving variational GNE of the time-varying game.
\item[3)] A decentralized bandit online algorithm is further devised for the studied online game when only the values of local cost functions and nonlinear constraint functions are available after all the players' decisions are made.
%Different from distributed bandit online algorithms for solving optimization problems \cite{yi2020distributed}, where two or more-point samplings can be used, our algorithm is more challenge to estimate the gradients of the cost functions and nonlinear constraint functions by using one-point stochastic approximation.
%\item[3)] Mirror descent is utilized in designing the algorithm of this paper, which is more applicable and general than projection-based algorithms \cite{lu2020online}, as the mirror descent algorithm subsumes a few interesting cases, such as the online projection-based algorithm \cite{lu2020online} and the entropic descent algorithm \cite{beck2003mirror}.
\end{itemize}

The rest of this paper is structured as follows. In Section \ref{section2}, some preliminaries and the problem formulation are introduced. Section \ref{section3} presents a decentralized learning algorithm for the studied online game in the setup of full information feedback, which is proved to be no-regret. The equilibrium convergence property of the designed algorithm is presented in Section \ref{section4}. In Section \ref{section5}, a bandit feedback case is discussed.
A numerical example is presented to show the effectiveness of the proposed algorithms in Section \ref{section6}. Section \ref{section7} makes a brief conclusion.

{\em Notations.} The symbols $\mathbb{R}$, $\mathbb{R}^m$, and $\mathbb{R}^{n\times m}$ represent the sets of real numbers, $m$-dimensional real column vectors, and $n\times m$ real matrices, respectively. Let $\mathbb{R}^m_+$ be the set of $m$-dimensional nonnegative vectors.
%$\mathbb{S}^m$ denotes the sphere centered at the origin in $\mathbb{R}^m$, while $\mathbb{B}^m$ is the unit ball centered at the origin in $\mathbb{R}^m$.
The symbol ${\bf 1}_m$ (resp. ${\bf 0}_m$) represents an $m$-dimensional vector, whose entries are 1 (resp. 0). For a vector or matrix $A$, the transpose of $A$ is denoted by $A^\top$.  The identity matrix of dimension $m$ is denoted by $I_m$. For an integer $m>0$, let $[m]:=\{1,2,\ldots,m\}$. Let $col(y_1,\ldots,y_m):=(y_1^{\top},\ldots,y_m^{\top})^{\top}$. $P\otimes Q$ denotes the Kronecker product of matrices $P$ and $Q$. For a vector $v\in\mathbb{R}^m$, $[v]_+$ is the projection of $v$ onto $\mathbb{R}^m_+$. $\langle x,y\rangle$ is the standard inner product of $x\in\mathbb{R}^m$ and $y\in\mathbb{R}^m$. For two vectors $x,y\in\mathbb{R}^m$, the symbol $x\leq y$ means that each entry of $x-y$ is nonpositive, while for two real symmetric matrices $W,P\in\mathbb{R}^{m\times m}$, $W\succeq P$ and $W\succ P$ mean that $W-P$ is positive semi-definite and positive definite, respectively. Given two functions $h_1(\cdot)$ and $h_2(\cdot)$, the notations $h_1=\mathcal{O}(h_2)$ and $h_1={\bf o}(T)$ mean that there exists a positive constant $C>0$ such that $|h_1(x)|\leq Ch_2(x)$ and $\lim\limits_{T\to\infty}\frac{h_1}{T}=0$ for all $x$ in the domain, respectively. A mapping $F:X\to Y$, where $X,Y\subseteq\mathbb{R}^n$, is $\mu$-strongly monotone ($\mu>0$), if for any $x_1,x_2\in X$, there holds $\langle F(x_1)-F(x_2),x_1-x_2\rangle\geq\mu\|x_1-x_2\|^2$. ${\bf E}[\cdot]$ is the expectation operator.

%${\rm diag}\{a_1,a_2,\ldots,a_n\}$ represents a diagonal matrix with $a_i$, $i\in[n]$, on its diagonal. For a vector $v$, we use ${\rm diag}(v)$ to represent the diagonal matrix with the vector $v$ on its diagonal.

\section{Preliminaries}\label{section2}
\subsection{Graph Theory}\label{subII.A}
 Let ${\mathcal G}=({\mathcal V},{\mathcal E},A)$ be an undirected graph, where the vertex set is ${\mathcal V}=[N]$, the edge set is ${\mathcal E}\subseteq{\mathcal V}\times{\mathcal V}$ and the weighted adjacency matrix is $A=(a_{ij})\in\mathbb{R}^{N\times N}$. For any $i,j\in[N]$ and $i\neq j$, $a_{ij}>0$ if $(j,i)\in{\mathcal E}$ and $a_{ij}=0$ otherwise. In this paper, it is assumed that $a_{ii}>0$ for all $i\in[N]$. $j$ is called a neighbor of $i$ if $(j,i)\in\mathcal{E}$. Denote ${\mathcal N}_i:=\{j:~(j,i)\in{\mathcal E}\}$. A path from node $i_1$ to node $i_l$ is composed of a sequence of edges $(i_h,i_{h+1})$, $h=1,2,\ldots,l-1$. It is said that an undirected graph ${\mathcal G}$ is connected if there exists a path from node $i$ to node $j$ for any two vertices $i,j$.
For communication graph ${\mathcal G}$, the following standard assumptions are imposed in this paper.
 \begin{assumption}\label{assump1}
The undirected graph ${\mathcal G}$ is connected and the adjacency matrix $A$ satisfies that $A^{\top}=A$ and $A{\bf1}_N={\bf1}_N$.
 \end{assumption}

%Let $A_i^-$ be a matrix in $\mathbb{R}^{(N-1)\times(N-1)}$, obtained by deleting the $i$th row and $i$th column of $A$. By Assumption \ref{assump1} and Lemma 3 in \cite{hong2006tracking}, one has that $I_{N-1}-A_i^-\succ0$. Therefore, all the eigenvalues of $A_i^-$ are less than 1. By the Gershgorin circle theorem, it can be easily derived that {\color{blue}each eigenvalue of $A_i^-$ is larger than $-1$. Thus, $\|A_i^-\|<1$ for every $i\in[N]$. Denote $\sigma:=\max_{i\in[N]}\|A_i^-\|\in(0,1)$.
Denote $\sigma:=\|A-{\bf1}_N{\bf1}_N^{\top}/N\|$. Then $0<\sigma<1$ under Assumption \ref{assump1} \cite{li2020}.

\subsection{Problem Formulation}
%Denote by $\Gamma(\mathcal{V},\Omega,J)$ a noncooperative game with $N$ players, where $\mathcal{V}:=[N]$ is the set of players, $\Omega:=\Omega_1\times\cdots\times\Omega_N$ represents the strategy set of players with $\Omega_i\subseteq\mathbb{R}^{n_i}$ being the private action set of player $i$, and $J=(J_1,\ldots,J_N)$ is the cost function with $J_i$ being the cost function of player $i$. Denote by $x=col(x_1,\ldots,x_N)$ the joint action, where $x_i$ is the action of player $i$, $i\in[N]$. Denote by $x_{-i}:=col(x_1,\ldots,x_{i-1},x_{i+1},\ldots,x_N)$ the joint action of all the players except $i$. For a game $\Gamma(\mathcal{V},\Omega,J)$, a strategy profile $x^*=(x_1^*,\ldots,x_N^*)\in\Omega$ is called an NE if
%\begin{align}
%J_i(x_i^*,x_{-i}^*)\leq J_i(x_i,x_{-i}^*), ~\forall x_i\in\Omega_i, ~i\in[N].
%\end{align}
%Moreover, if $\Omega_i$ depends on other players' actions, then the NE $x^*$ is termed a GNE.

This paper studies online game $\Gamma_t:=\Gamma(\mathcal{V},\Omega_t,J_t)$ with $N$ players, where $\mathcal{V}:=[N]$ is the set of players, $\Omega_t=\Omega_{0,t}\cap(\Omega_{1}\times\cdots\times\Omega_{N})$ represents the time-varying strategy set of players with $\Omega_{0,t}$ being the shared convex constraint $\Omega_{0,t}:=\{x=col(x_1,\ldots,x_N)\in\mathbb{R}^n\mid x_i\in\mathbb{R}^{n_i}, i\in[N], g_t(x):=\sum_{i=1}^Ng_{i,t}(x_i)\leq{\bf 0}_m\}$ and $\Omega_{i}$ being the private action set constraint of player $i\in[N]$, and $J_t=(J_{1,t},\ldots,J_{N,t})$ is the time-varying cost function with $J_{i,t}$ being the private cost function of player $i$. Here, $n:=\sum_{i=1}^Nn_i$ and $g_{i,t}:\mathbb{R}^{n_i}\to\mathbb{R}^m$, $i\in[N]$. Denote by $x_t=col(x_{1,t},\ldots,x_{N,t})$ the joint action at time $t$, where $x_{i,t}$ is the action of player $i$ at time $t$, $i\in[N]$. Denote by $x_{-i,t}:=col(x_{1,t},\ldots,x_{i-1,t},x_{i+1,t},\ldots,x_{N,t})$ the joint action at time $t$ of all the players except $i$. For this time-varying game, it is impossible for every player to pre-compute a GNE $(x_{i,t}^*,x_{-i,t}^*)$ of $\Gamma_t$, satisfying
\begin{align}
J_{i,t}(x_{i,t}^*,x_{-i,t}^*)\leq J_{i,t}(x_i,x_{-i,t}^*), ~\forall x_i\in\Omega_{i,t}(x_{-i,t}^*),
\end{align}
where $\Omega_{i,t}(x_{-i,t}^*):=\{x_i\in\Omega_i\mid (x_i,x_{-i,t}^*)\in\Omega_t\}$. Thus, a new benchmark needs to be proposed to depict the performance of decentralized learning algorithms, such as regret, which is commonly regarded as a concise and meaningful benchmark for quantifying the ability of an online algorithm for decision-making in the presence of uncertainties and unpredictabilities \cite{cesa2006prediction}. In general, the (static) regret for player $i$ is defined as follows:
 \begin{align}\label{equ7}
 Reg_i(T):=\sum_{t=1}^T[J_{i,t}(x_{i,t},x_{-i,t})-J_{i,t}(\tilde{x}_i,x_{-i,t})],
 \end{align}
 where $T$ is the learning time and $$\tilde{x}_i:=\arg\min\limits_{x_i\in\cap_{t=1}^T\Omega_{i,t}(x_{-i,t})}\sum_{t=1}^TJ_{i,t}(x_i,x_{-i,t}).$$
 Intuitively, the regret captures the cumulative payoff difference between a player's actual policy and the best action in hindsight.

Since the strategies of players should satisfy a coupled nonlinear constraint, a notion of regret associated with the constraint is also required. Specifically, for a decision sequence $\{x_{1},\ldots,x_{T}\}$, the constraint violation measure is given as
 \begin{align}\label{equ8}
 R_g(T):=\left\|\left[\sum\limits_{t=1}^Tg_t(x_t)\right]_+\right\|.
 \end{align}

It is said that an online algorithm is no-regret if all the regrets of players in (\ref{equ7}) and the accumulation of constraint violations in (\ref{equ8}) increase sublinearly, i.e.,
$Reg_i(T)={\bf o}(T)$ and $R_g(T)={\bf o}(T)$.

In this paper, the objective is to devise no-regret algorithms for the studied online game and to discuss the equilibrium convergence properties of the designed algorithms.

To end this section, a blanket assumption on game $\Gamma_t$ is presented, which is also made in \cite{salehisadaghiani2016distributed,lu2020online}. For each player $i\in[N]$, let $g_{ij,t}$ denote the $j$th component of $g_{i,t}$, i.e., $g_{i,t}=col(g_{i1,t},\ldots,g_{im,t})$.

\begin{assumption}\label{assump2}
For each $i\in[N]$, the set $\Omega_i$ is compact and convex. $J_{i,t}(\cdot,x_{-i}):\Omega_i\to\mathbb{R}$ and $g_{ij,t}(\cdot)$ are convex. $\{J_{i,t}\}$ and $\{g_{i,t}\}$ are uniformly bounded. Then there exists $L>0$ such that for any $x_i\in\Omega_i$, $i\in[N]$,
\begin{align}
	\|x_i\|<L, ~\|J_{i,t}(x)\|\leq L, ~\|g_{i,t}(x_i)\|\leq L.\label{equ2}
\end{align}
$\{\nabla_iJ_{i,t}\}$ and $\{\nabla g_{ij,t}\}$ exist and are uniformly bounded, that is, there exists $M>0$ such that for any $x_i\in\Omega_i$, $i\in[N]$,
\begin{align}
	\|\nabla_iJ_{i,t}(x)\|\leq M,~\|\nabla g_{i,t}(x_i)\|\leq M,\label{equ3}
\end{align}
where $\nabla g_{i,t}(x_i):=(\nabla g_{i1,t}(x_i),\ldots,\nabla g_{im,t}(x_i))$ and $\nabla_iJ_{i,t}(x_i,x_{-i}):=\frac{\partial J_{i,t}(x_i,x_{-i})}{\partial x_i}$.
Moreover, the constraint set $\Omega_t$ is assumed to be non-empty and Slater's constraint qualification is satisfied.
\end{assumption}

\subsection{Bregman Divergence}
For each player $i\in[N]$, the Bregman divergence $D_{\phi_i}(\xi,\zeta)$ of two points $\xi,\zeta\in\Omega_i\subseteq\mathbb{R}^{n_i}$ is defined as \cite{bregman1967relaxation}
\begin{align}
D_{\phi_i}(\xi,\zeta):=\phi_i(\xi)-\phi_i(\zeta)-\langle\nabla\phi_i(\zeta),\xi-\zeta\rangle,
\end{align}
where $\phi_i:\Omega_i\to\mathbb{R}$ is differentiable and $\mu_i$-strongly convex for some constant $\mu_i>0$, i.e., $\phi_i(\xi)\geq\phi_i(\zeta)+\langle\nabla\phi_i(\zeta),\xi-\zeta\rangle+\frac{\mu_i}{2}\|\xi-\zeta\|^2.$ Thus, it can be easily derived that $D_{\phi_i}(\cdot,\zeta)$ is $\mu_0$-strongly convex for all $i\in[N]$ with $\mu_0:=\min\{\mu_1,\ldots,\mu_N\}$, i.e.,
\begin{align}\label{equ10}
D_{\phi_i}(\xi,\zeta)\geq\frac{\mu_0}{2}\|\xi-\zeta\|^2,
\end{align}
and the generalized triangle inequality is satisfied, i.e.,
\begin{align}
&\langle \xi-\zeta, \nabla\phi_i(\zeta)-\nabla\phi_i(\theta)\rangle\nonumber\\
&=D_{\phi_i}(\xi,\theta)-D_{\phi_i}(\xi,\zeta)-D_{\phi_i}(\zeta,\theta).\label{equ11}
\end{align}

Bregman divergence is widely applied in machine learning and game theory, generalizing the standard Euclidean distance and the generalized Kullback-Leibler divergence. Specifically, when $\phi_i(\theta)=\|\theta\|^2$, Bregman divergence $D_{\phi_i}(\theta,\vartheta)$ amounts to $D_{\phi_i}(\theta,\vartheta)=\|\theta-\vartheta\|^2$. If $\phi_i(\xi)=\sum_{j=1}^{n_i}\xi_j\log \xi_j$, then the Bregman divergence becomes the generalized Kullback-Leibler divergence $D_{\phi_i}(\xi,\zeta)=\sum_{j=1}^{n_i}\xi_j\log\frac{\xi_j}{\zeta_j}-\sum_{j=1}^{n_i}\xi_j+\sum_{j=1}^{n_i}\zeta_j$.
A mild assumption on Bregman divergence is given below.
\begin{assumption}\label{assump3}
For any $i\in[N]$ and $\xi,\zeta\in\Omega_{i}$, $D_{\phi_i}(\xi,\zeta)$ is Lipschitz with respect to the first variable $\xi\in\Omega_i$, i.e., one can find a positive number $K$ such that for any $\xi_1,\xi_2\in\Omega_{i}$,
\begin{align}
|D_{\phi_i}(\xi_1,\zeta)-D_{\phi_i}(\xi_2,\zeta)|\leq K\|\xi_1-\xi_2\|.
\end{align}
\end{assumption}
%\begin{assumption}\label{assump6}
%%For any $i\in[N]$, $\nabla\phi_i(x)$ is bounded for any $x\in\Omega_i$, that is, there exists a positive constant $K$ such that $\|\nabla\phi_i(x)\|\leq K$.
%For any $i\in[N]$ and $\xi\in\Omega_{i}$, $D_{\phi_i}(\xi,\cdot):\Omega_{i}\to\mathbb{R}$ is convex, i.e., for any $a\in[0,1]$,
%\begin{align}
%&D_{\phi_i}(\xi,a\zeta_1+(1-a)\zeta_2)\nonumber\\
%&\leq aD_{\phi_i}(\xi,\zeta_1)+(1-a)D_{\phi_i}(\xi,\zeta_2),~\forall \zeta_1,\zeta_2\in\Omega_{i}.
%\end{align}
%\end{assumption}

Assumption \ref{assump3} can be satisfied when $\phi_i(\xi)$ is Lipschitz on $\Omega_{i}$, which implies that for any $\xi,\zeta\in\Omega_i$, $D_{\phi_i}(\xi,\zeta)=|D_{\phi_i}(\xi,\zeta)-D_{\phi_i}(\zeta,\zeta)|\leq K\|\xi-\zeta\|$.
%Assumption \ref{assump6} is crucial to derive the main results in this paper and a sufficient condition given in \cite{bauschke2001joint} for guaranteeing Assumption \ref{assump6} is that $\phi_i(\xi)$ is thrice continuously differentiable and satisfies
%%$H_{\phi_i}(\xi)\succeq 0$, $H_{\phi_{i}}(\xi)+\nabla H_{\phi_i}(\xi)(\xi-\zeta)\succeq0$, where $\xi,\zeta\in\Omega_{i}$ and $H_{\phi_i}$ represents the Hessian matrix of $\phi_i$.

\section{Regret Minimization}\label{section3}
At each time $t$, player $i$ in the online game $\Gamma_t$ only receives the values of  $J_{i,t}(x_{i,t},x_{-i,t})$, $g_{i,t}(x_{i,t})$, $\nabla_iJ_{i,t}(x_{i,t},x_{-i,t})$ and $\nabla g_i(x_{i,t})$, denoted as $R_{i,t}$, $C_{i,t}$, $V_{i,t}$ and $G_{i,t}$, respectively. In the following, a decentralized online algorithm based on mirror descent and primal-dual methods will be devised for online game $\Gamma_t$.

For each $i\in[N]$, define an augmented Lagrangian function at time $t$ as
$
{\cal L}_{i,t}(x_{i,t},\lambda_t;x_{-i,t}):=J_{i,t}(x_{i,t},x_{-i,t})+\lambda^{\top}_tg_t(x_t)-\frac{\beta_{t}}{2}\|\lambda_t\|^2,
$ where $\lambda_t\in\mathbb{R}^m_+$ is the Lagrange multiplier or the dual variable, $\gamma_t>0$ is a stepsize, and $\beta_t>0$ is the regularization parameter.
Inspired by the dynamic mirror descent for online optimization \cite{hall2015online}, a semi-decentralized primal-dual dynamic mirror descent can be designed as
\begin{align}
x_{i,t+1}&=\arg\min\limits_{x\in\Omega_i}\{\alpha_{t}\langle x,V_{i,t}+G_{i,t}\lambda_t\rangle\nonumber\\
&~~~~~~~~~~~~~~~+D_{\phi_i}(x,x_{i,t})\},\label{equ14}\\
\lambda_{t+1}&=\left[\lambda_t+\gamma_t(g_t(x_t)-\beta_t\lambda_t)\right]_+,\label{equ15}
\end{align}
where $\alpha_t>0$ and $\gamma_t\in(0,1)$ are time-varying stepsizes utilized in the primal and dual iterations. The main drawbacks of this algorithm are that each player should know the common Lagrange multiplier $\lambda_t$ and the global nonlinear constraint function $g_t(x_t)$, which require a central coordinator to bidirectionally communicate with all the players.
Motivated by the algorithm proposed in \cite{neely2017online}, by modifying (\ref{equ14}) and (\ref{equ15}), a decentralized online primal-dual dynamic mirror descent algorithm is designed as in Algorithm \ref{alg1} for online game $\Gamma_t$.
%%The parameters $\alpha_t,\beta_t,\gamma_t,\delta_t$ play a key role in deriving the main result.
%To execute Algorithm \ref{alg1}, at each time slot $t$, every player $i$ needs to know $\nabla_iJ_{i,t}({\bf x}_{i,t})$, $g_{i,t}(x_{i,t})$ and $\nabla g_{i,t}(x_{i,t})$ rather than the full information of $J_{i,t}$ and $g_{i,t}$, which is similar to most online algorithms for optimization and games \cite{mahdavi2012trading,jenatton2016adaptive,yuan2018online,lu2020online,li2020distributed,li2020distributed_online}.

\begin{algorithm}[!htbp]\caption{Decentralized Online Primal-Dual Dynamic Mirror Descent}\label{alg1}
Each player $i$ maintains vector variables $x_{i,t}\in\mathbb{R}^{n_i}$ and $\lambda_{i,t}\in\mathbb{R}^{m}$ at iteration $t\in[T]$.

{\bf Initialization:} For any $i\in[N]$, initialize $x_{i,1}\in\Omega_i$ arbitrarily and $\lambda_{i,1}={\bf0}_m$.

{\bf Iteration:} For $t\geq 1$, every player $i$ processes the following update:
\begin{align}
x_{i,t+1}&=\arg\min\limits_{x\in\Omega_i}\{\alpha_t\langle x,V_{i,t}+G_{i,t}\tilde{\lambda}_{i,t}\rangle\nonumber\\
&~~~~~~~~~~~~+D_{\phi_i}(x,x_{i,t})\},\label{equ16}\\
\lambda_{i,t+1}&=[\tilde{\lambda}_{i,t}+\gamma_t(C_{i,t}-\beta_t\tilde{\lambda}_{i,t})]_+,\label{equ17}
\end{align}
where $\tilde{\lambda}_{i,t}:=\sum_{j=1}^Na_{ij}\lambda_{j,t}$,
%$b_{i,t+1}:=\nabla g_{i,t}(x_{i,t})(\tilde{x}_{i,t+1}-x_{i,t})+g_{i,t}(x_{i,t})$,
$a_{ij}$ is the $(i,j)$th element of $A$, and $\alpha_t,\beta_t,\gamma_t$, satisfying $\alpha_0=\beta_0=\gamma_0=1$, are the stepsizes to be determined.
\end{algorithm}

%It should be noted that Algorithm \ref{alg1} subsumes a few interesting cases since Bregman divergence generalizes the standard Euclidean distance and the generalized Kullback-Leibler divergence, which are widely applied in machine learning and game theory. For example, let $\phi_i(\xi)=\frac{1}{2}\|\xi\|^2$, then the associated Bregman divergence is $D_{\phi_i}(\xi,\zeta)=\frac{1}{2}\|\xi-\zeta\|^2$, thus leading to that Algorithm \ref{alg1} reduces to a variant of the decentralized online projection-based algorithm in \cite{lu2020online}. As another example, let $\phi_i(\xi)=\sum_{i=1}^{n_i}[\xi]_i\ln[\xi]_i$ with $\xi\in\Omega_i:=\{\xi\in\mathbb{R}^{n_i}_+\mid \sum_{i=1}^{n_i}[\xi]_i=1\}$, where $[\xi]_i$ represents the $i$th element of $\xi$, then the associated Bregman divergence is the Kullback-Leibler divergence $D_{\phi_i}(\xi,\zeta)=\sum_{i=1}^{n_i}[\xi]_i\ln\frac{[\xi]_i}{[\zeta]_i}$. In this case, the iteration (\ref{equ16}) reduces to a revised version of the distributed entropic descent algorithm in \cite{beck2003mirror}, while the projection step in the conventional projected gradient algorithm does not have a closed-form solution.

In what follows, some necessary lemmas are presented.
\begin{lemma}\label{lemma1}
If Assumptions \ref{assump1} and \ref{assump2} hold, then for any $i\in[N]$ and $t\in[T]$, $\lambda_{i,t}$ and $\tilde{\lambda}_{i,t}$ generated by Algorithm \ref{alg1} satisfy
\begin{align}
\|\lambda_{i,t}\|&\leq\frac{L}{\beta_t},\label{equ18}\\
\|\tilde{\lambda}_{i,t}\|&\leq\frac{L}{\beta_t},\label{equ19}\\
\|\tilde{\lambda}_{i,t}-\overline{\lambda}_t\|&\leq2\sqrt{N}L\sum\limits_{s=0}^{t-1}\sigma^{s}\gamma_{t-1-s},\label{equ20}\end{align}
\begin{align}
\frac{\Xi_{t+1}}{2\gamma_t}&\leq2NL^2\gamma_t+(\overline{\lambda}_t-\lambda)^{\top}C_t+\frac{N\beta_t}{2}\|\lambda\|^2\nonumber\\
&~~~+c(t),\label{equ21}
\end{align}
where $\lambda\in\mathbb{R}^m_+$, $\overline{\lambda}_t:=\frac{1}{N}\sum_{i=1}^N\lambda_{i,t}$, $C_t:=\sum_{i=1}^NC_{i,t}$, $c(t):=2N\sqrt{N}L^2\sum\limits_{s=0}^{t-1}\sigma^{s}\gamma_{t-1-s}$, and
$
\Xi_{t+1}:=\sum_{i=1}^N\left[\|\lambda_{i,t+1}-\lambda\|^2-(1-\beta_t\gamma_t)\|\lambda_{i,t}-\lambda\|^2\right].
$
\end{lemma}

{\em Proof:} See Appendix \ref{A}. \hfill$\blacksquare$

\begin{lemma}\label{lemma2}
Under Assumptions \ref{assump1} and \ref{assump2}, for any $i\in[N]$ and $t\in[T]$, $x_{i,t}$ generated by Algorithm \ref{alg1} satisfies
\begin{align}
\langle x_{i,t}-x_i,V_{i,t}\rangle&\leq\frac{M^2}{\mu_0}\alpha_t+4\sqrt{N}L^2\sum_{s=0}^{t-1}\sigma^s\gamma_{t-1-s}\nonumber\\
&~+\frac{M^2L^2}{\mu_0}\frac{\alpha_t}{\beta_t^2}+\overline{\lambda}_t^{\top}(g_{i,t}(x_i)-g_{i,t}(x_{i,t}))\nonumber\\
&~+\frac{1}{\alpha_t}[D_{\phi_i}(x_i,x_{i,t})-D_{\phi_i}(x_i,x_{i,t+1})]\label{equ29}
\end{align}
for any $x_i\in\Omega_i$.
\end{lemma}

{\em Proof:} See Appendix \ref{B}. \hfill$\blacksquare$

\begin{lemma}\label{lemma3}
Under Assumptions \ref{assump1}--\ref{assump3}, for any $i\in[N]$, the regret (\ref{equ7}) generated by Algorithm \ref{alg1} is bounded by
\begin{align}
Reg_i(T)&\leq B_1(T)+B_2(T),\label{equ35}
\end{align} and the accumulated constraint violation satisfies
 \begin{align}
R_g(T)\leq \sqrt{B_1(T)B_3(T)+B_3(T)B_4(T)}, \label{equ36}
\end{align}
where
\begin{align*}
B_1(T)&:=\frac{2LK}{\alpha_{T+1}}+\Big(\frac{2(N+2)\sqrt{N}L^2}{1-\sigma}+2NL^2\Big)\sum_{t=1}^T\gamma_{t-1}\\
&~~~~+\frac{M^2}{\mu_0}\sum_{t=1}^T\alpha_{t}+\frac{L^2M^2}{\mu_0}\sum_{t=1}^{T}\frac{\alpha_t}{\beta_t^2},\nonumber\\
B_2(T)&:=\frac{1}{2}\sum_{t=1}^T\sum\limits_{i=1}^N\left(\frac{1}{\gamma_t}-\frac{1}{\gamma_{t-1}}-\beta_t\right)\|\lambda_{i,t}\|^2,\nonumber\\
B_3(T)&:=2N\left(1+\sum_{t=1}^T\beta_t\right),\nonumber\\
B_4(T)&:=\frac{1}{2}\sum_{t=1}^T\sum\limits_{i=1}^N\left(\frac{1}{\gamma_t}-\frac{1}{\gamma_{t-1}}-\beta_t\right)\|\lambda_{i,t}-\lambda_c\|^2,
\end{align*}
and $\lambda_c:=\frac{2\left[\sum_{t=1}^Tg_t(x_t)\right]_+}{B_3(T)}$.
\end{lemma}

{\em Proof:} See Appendix \ref{C}. \hfill$\blacksquare$

It is now ready to give the sublinear bounds on the regrets and constraint violation for Algorithm \ref{alg1}.
\begin{theorem}\label{thm1}
If Assumptions \ref{assump1}--\ref{assump3} hold, then for each $i\in[N]$ and the sequence $\{x_{i,1},\ldots,x_{i,T}\}$ generated by Algorithm \ref{alg1} with
$$\alpha_t=\frac{1}{t^{a_1}},\beta_t=\frac{1}{t^{a_2}},\gamma_t=\frac{1}{t^{1-a_2}},$$ where $\alpha_0=\beta_0=\gamma_0=1$, $0<a_1<1$ and $a_1>2a_2>0$, there hold
\begin{align}
Reg_i(T)&=\mathcal{O}(T^{\max\{a_1,1-a_1+2a_2\}}),\\
R_g(T)&=\mathcal{O}(T^{\max\{\frac{1}{2}+\frac{a_1}{2}-\frac{a_2}{2},1-\frac{a_1}{2}+\frac{a_2}{2}\}}).\label{}
\end{align}
%where $C:=M\sqrt{{C_0}}$ and ${C_0}:=\max\{\frac{1}{\mu}(2NLK+C_1+\frac{C_2}{1-a_1}),\frac{\sqrt{N}K}{\mu}\}$.
\end{theorem}

{\em Proof:}
For the selected parameters $\beta_t$ and $\gamma_t$, it is easy to verify that $\frac{1}{\gamma_t}-\frac{1}{\gamma_{t-1}}-\beta_t\leq0$.
In addition, for any constant $0<a\neq1$ and positive integer $T$, it holds that
\begin{align}\label{equ50}
\sum\limits_{t=1}^T\frac{1}{t^a}\leq1+\int_{1}^T\frac{1}{t^{a}}dt\leq1+\frac{T^{1-a}-1}{1-a}\leq\frac{T^{1-a}}{1-a}.
\end{align}
Then,
\begin{align}
\sum_{t=1}^T\alpha_t&\leq\frac{T^{1-a_1}}{1-a_1}.
\end{align}
Thus,
\begin{align}
B_1(T)&=\mathcal{O}(T^{a_1}+T^{a_2}+T^{1-a_1}+T^{1-a_1+2a_2})\nonumber\\
&=\mathcal{O}(T^{a_1}+T^{1-a_1+2a_2}),\\
B_3(T)&=\mathcal{O}(T^{1-a_2}).\label{equ53}
\end{align}
Subsequently, one can obtain from Lemma \ref{lemma3} that
\begin{align*}
Reg_i(T)&=\mathcal{O}(T^{\max\{a_1,1-a_1+2a_2\}}),
\end{align*}
and
\begin{align*}
R_g(T)=\mathcal{O}(T^{\max\{\frac{1}{2}+\frac{a_1}{2}-\frac{a_2}{2},1-\frac{a_1}{2}+\frac{a_2}{2}\}}).
\end{align*}
The result is thus proved.
\hfill$\blacksquare$

\begin{corollary}
 Under Assumptions \ref{assump1}--\ref{assump3}, Algorithm \ref{alg1} achieves sublinearly bounded regrets and constraint violation if $0<2a_2<a_1<1$.
\end{corollary}

{\em Proof:}
By the selection of $\alpha_1$ and $\alpha_2$, one has that $1-a_1+2a_2$, $\frac{1}{2}+\frac{a_1}{2}-\frac{a_2}{2}$ and $1-\frac{a_1}{2}+\frac{a_2}{2}$ are in $(0,1)$. Therefore, $Reg_i(T)={\bf o}(T)$ and $R_g(T)={\bf o}(T)$.
\hfill$\blacksquare$

\begin{corollary}\label{corollary2}
 Under Assumptions \ref{assump1}--\ref{assump3}, Algorithm \ref{alg1} can achieve sublinear bounds on the regrets and constraint violation as
\begin{align*}
Reg_i(T)&=\mathcal{O}(T^{\frac{1}{2}+a_2}),\\
R_g(T)&=\mathcal{O}(T^{\frac{3}{4}}),
\end{align*} where $a_2\in(0,0.5)$.
\end{corollary}

{\em Proof:}
Let $a_1=1-a_1+2a_2$, then it is derived that $a_1=1/2+a_2$ and
$\frac{1}{2}+\frac{a_1}{2}-\frac{a_2}{2}=1-\frac{a_1}{2}+\frac{a_2}{2}=3/4$. The result is thus proved.
\hfill$\blacksquare$

\begin{remark}
The regret bound in this paper is almost the same as the well-known order-optimal bound $\mathcal{O}(\sqrt{T})$ \cite{abernethyoptimal} when the constant $\alpha_2$ is chosen small enough.  In addition, the bound of the accumulated constraint violation is also sublinear with the bound $\mathcal{O}(T^{3/4})$. To our best knowledge, the obtained results here are the first ones for online game with time-varying coupled inequality constraints.
\end{remark}
\begin{remark}
The feedback signal after choosing an action, such as the values $V_{i,t}$ and $G_{i,t}$, is one of the most important ingredients in designing the learning scheme. In general, the exact gradients' information is usually hard to grab because of stochastic factors and unknown structures of the payoff and constraint functions. Specifically, for an action profile $x_t=col(x_{1,t},\ldots,x_{N,t})$, assume that the feedback signal is the noisy gradients, that is,
\begin{align*}
V_{i,t}&=\nabla_iJ_{i,t}(x_t)+v_{i,t},\\
G_{i,t}&=\nabla g_{i,t}(x_{i,t})+w_{i,t},
\end{align*}
where $v_{i,t}$ and $w_{i,t}$ are random observation noises capturing all uncertainties and stochasticities. In this setting, the learning policy in Algorithm \ref{alg1} is still applicable to generate a sequence of actions that ensures the sublinear bounds on the expected regrets and constraint violation by properly selecting stepsizes and under some mild conditions on the means and variances of noise variables $v_{i,t}$ and $w_{i,t}$.
\end{remark}

\section{Tracking Nash Equilibria}\label{section4}
Another important issue is whether the learning policy, Algorithm \ref{alg1}, ensures that the actions of players converge or track the GNEs for the studied online game.  As pointed out in \cite{duvocelle2022multiagent} that establishing a blanket casual link between no-regret play and convergence to NEs is impossible because of unilateral and coupled relationship among players, analyzing the equilibrium convergence properties of the designed learning algorithm is much more challenging and difficult especially under nonlinear coupled constraints. In this section, it is assumed that online game $\Gamma_t$ stabilizes to a $\mu$-strongly monotone game $\Gamma:=\Gamma(\mathcal{V},\Omega,J)$, where $\Omega:=\Omega_0\cap(\Omega_1\times\cdots\times\Omega_N)$ with $\Omega_0:=\{x=col(x_1,\ldots,x_N)\in\mathbb{R}^n\mid x_i\in\mathbb{R}^{n_i}, i\in[N], g(x):=\sum_{i=1}^Ng_{i}(x_i)\leq{\bf 0}_m\}$ and $J=(J_1,\ldots,J_N)$ with $J_i(x_i,x_{-i})$ being convex and differentiable with respect to $x_i$. Here, this stabilization is formally defined as
\begin{align}
	&\lim_{t\to\infty}{H}_{i,t}=0,~\forall i\in[N],\\
	&\lim_{t\to\infty}{K}_{t}=0,
\end{align}
where $H_{i,t}:=\max\limits_{x\in\Omega_1\times\cdots\times\Omega_N}\|\nabla_iJ_{i,t}(x)-\nabla_iJ_i(x)\|$ and $K_{t}:=\max\limits_{x\in\Omega_1\times\cdots\times\Omega_N}\|g_t(x)-g(x)\|$.

It should be noted that the convergence is defined in terms of the payoff gradients rather than the payoff functions mainly due to that a GNE is a solution to a variational inequality only involving the payoff gradients of players.

For the $\mu$-strongly monotone game limit $\Gamma$, one has for any $x,y\in\Omega_{1}\times\cdots\times\Omega_{N}$,
\begin{align}
	(F(x)-F(y))^{\top}(x-y)\geq\mu\|x-y\|^2,
\end{align}
where  $F(x):=col(\nabla_1J_1(x_1,x_{-1}),\ldots,\nabla_NJ_N(x_N,x_{-N}))$ is called the {\em pseudo-gradient mapping} of game $\Gamma$. Note that the cost functions are convex and differentiable. Then it can be obtained from Theorem 3.9 in \cite{facchinei2009nash} that at any time $t$, a solution to the following variational inequality:
\begin{align}\label{e53}
	F^{\top}(x^*)(x-x^*)\geq 0, ~\text{for all}~x\in\Omega,
\end{align}
is a GNE of $\Gamma$, and this GNE $x^*$ is also called a variational GNE. In addition, the inequality (\ref{e53}) has a unique solution under the strong monotonicity of $F$. Therefore, the existence and uniqueness of the variational GNE can be guaranteed under the convex payoff functions and strongly monotone pseudo-gradient mapping. It is noticed that finding all GNEs is very difficult even if the game is offline. Accordingly, we will focus on tracking the unique variational GNE as done in \cite{liang2017distributed,pavel2020distributed} since the unique variational GNE enjoys good stability and has no price discrimination from the perspective of economics \cite{kulkarni2012variational}.

To proceed, a standard assumption is also needed.
\begin{assumption}\label{assump4}
The constraint set $\Omega$ is assumed to be non-empty and Slater's constraint qualification is satisfied.
\end{assumption}

Under Assumption \ref{assump4}, the optimal dual variable $\lambda^*\in\mathbb{R}^m_+$ to the Lagrangian function $\mathcal{L}_i(x_i,\lambda;x_{-i}):=J_i(x_i,x_{-i})+\lambda^{\top}g(x)$ associated to game $\Gamma$ is bounded (cf. \cite{nedic2009approximate}), that is, there exists a positive constant $\Lambda>0$ such that
\begin{align}
\|\lambda^*\|\leq\Lambda.\label{equ54}
\end{align}

Now, it is ready to present the result on the convergence of the play sequence generated by Algorithm \ref{thm1}.

\begin{theorem}\label{thm2}
	 Assume that online game $\Gamma_t$ converges to a strongly monotone game $\Gamma$. Under Assumptions \ref{assump1}--\ref{assump4}, if $\alpha_t$, $\beta_t$ and $\gamma_t$ are selected to satisfy
	 \begin{align}
	 	&\sum_{t=1}^{\infty}\alpha_t=\infty,\sum_{t=1}^{\infty}\alpha_t^2<\infty, \sum_{t=1}^{\infty}\frac{\alpha_t^2}{\beta_t^2}<\infty, \sum_{t=1}^{\infty}\alpha_t\beta_t<\infty,\nonumber\\
	 	&\sum_{t=1}^{\infty}\alpha_t\gamma_t<\infty,\sum_{t=1}^{\infty}\alpha_tc(t)<\infty,\sum_{t=1}^{\infty}\alpha_tH_{i,t}<\infty,\nonumber\\
	 	&\sum_{t=1}^{\infty}\frac{\alpha_t}{\beta_t}K_t<\infty,\frac{1}{\gamma_t}-\frac{1}{\gamma_{t-1}}-\beta_t\leq0,\label{equ55}
	 \end{align}
	  then the sequence $\{x_t\}$ generated by Algorithm \ref{alg1} converges to the variational GNE $x^*=col(x_1^*,\ldots,x_N^*)$ of game $\Gamma$.
\end{theorem}

{\em Proof:}
See Appendix \ref{D}.
\hfill$\blacksquare$

\begin{corollary}\label{corollary3}
	Assume that online game $\Gamma_t$ converges to a strongly monotone game $\Gamma$ with $H_{i,t}=\mathcal{O}(1/t^{p})$ for all $i\in[N]$ and $K_t=\mathcal{O}(1/t^q)$ for some $p,q>0$. Under Assumptions \ref{assump1}--\ref{assump4}, if $\alpha_t$, $\beta_t$ and $\gamma_t$ are selected as
$$\alpha_t=\frac{1}{t^{b_1}},~\beta_t=\frac{1}{t^{b_2}},~\gamma_t=\frac{1}{t^{1-b_2}},$$ where $\alpha_0=\beta_0=\gamma_0=1$, $0<b_2<0.5$, $0.5+b_2<b_1\leq1$, $b_1+b_2>1$, $b_1+p>1$ and $b_1-b_2+q>1$,
	then the sequence $\{x_t\}$ generated by Algorithm \ref{alg1} converges to the variational GNE $x^*=col(x_1^*,\ldots,x_N^*)$ of game $\Gamma$.
\end{corollary}

\begin{remark}
The selection of stepsizes $\alpha_t$, $\beta_t$ and $\gamma_t$ depends on the convergence orders of $H_{i,t}$ and $K_t$, which need to satisfy the conditions in (\ref{equ55}). For example, if the specific convergence orders of $H_{i,t}$ and $G_t$ are known as in Corollary \ref{corollary3}, then $b_2$ only needs to satisfy $b_2<\min\{0.5,q\}$ when choosing $b_1=1$.
\end{remark}

In Theorem \ref{thm2} and Corollary \ref{corollary3}, only the convergence of the generated play sequence is acquired, while the convergence rate is difficult to establish. Instead, the convergence rate of the averaged strategy sequence is derived in what follows.
\begin{theorem}\label{thm3}
	Assume that online game $\Gamma_t$ converges to a strongly monotone game $\Gamma$ with $H_{i,t}=\mathcal{O}(1/t^{p})$ for all $i\in[N]$ and $K_t=\mathcal{O}(1/t^q)$ for some $p,q>0$. Under Assumptions \ref{assump1}--\ref{assump4}, if $\alpha_t$, $\beta_t$ and $\gamma_t$ are selected as
	$$\alpha_t=\frac{1}{t^{b_1}},~\beta_t=\frac{1}{t^{b_2}},~\gamma_t=\frac{1}{t^{1-b_2}},$$
	where $\alpha_0=\beta_0=\gamma_0=1$, $0<2b_2<b_1<1$ and $b_2<q$,
	then the sequence $\{x_t\}$ generated by Algorithm \ref{alg1} satisfies
	\begin{align}
		\Big\|\overline{x}_T-x^*\Big\|^2=\mathcal{O}(T^{-\min\{1-b_1,b_1-2b_2,b_2,p,q-b_2\}}),\label{equ3-}
	\end{align}
where $\overline{x}_T:=\frac{1}{T}\sum_{t=1}^Tx_t$.
\end{theorem}

{\em Proof:}
	See Appendix \ref{E}.
\hfill$\blacksquare$

\begin{remark}
	From Theorem \ref{thm3}, it  can be seen that the convergence speed of the game sequence $\Gamma_t$ has an impact on the convergence rate of the generated averaged strategy $\overline{x}_T$. If $p,q$ are larger, then the convergence of $\overline{x}_T$  to $x^*$ is faster. However, when $p$ and $q$ are large enough, that is, game $\Gamma_t$ converges to the strongly monotone game $\Gamma$ fast enough,  or in other words, $\min\{p,q-b_2\}>\min\{1-b_1,b_1-2b_2,b_2\}$, then $\|\overline{x}_T-x^*\|^2=\mathcal{O}(T^{-\min\{1-b_1,b_1-2b_2,b_2\}})$, which implies that the convergence rate of $\overline{x}_T$ is independent of $p,q$.
\end{remark}

\section{Payoff-Based Learning}\label{section5}
%In many significant situations, the players may not know the actual structures of payoff functions and even not realize what the game scheme they participate is.
In this section, a payoff-based learning scheme, i.e., an online algorithm that only depends on payoff values after decisions are made, is devised, which is also called bandit feedback. First, the gradient estimator based on one-point bandit feedback is introduced. Then, by the analysis of the previous section, the convergence result for the payoff-based algorithm is presented.

The idea to estimate the payoff gradients relies on the so-called one-point stochastic approximation. Let $f:\mathbb{X}\to\mathbb{R}$ be a function, where $\mathbb{X}\subseteq\mathbb{R}^d$ is a convex set. To estimate the gradient $f(x)$ for some $x\in\mathbb{X}$, it suffices to sample $f$ at $x+\delta w$, where $\delta>0$ is a constant and $w\in\mathbb{R}^d$ is a vector taking values uniformly at random from $\{\pm{\bf e}_1,\pm{\bf e}_2,\ldots,\pm\bf{ e}_d\}$. Here, for each $i\in[N]$, ${\bf e}_i\in\mathbb{R}^d$ is a vector with its $i$th element being 1 and others 0. Then the estimate of $\nabla f(x)$ is
\begin{align*}
\nabla f(x)\approx\frac{d}{\delta}f(x+\delta w)w.
\end{align*}

Due to the constraint set $\mathbb{X}$, a problem that may arise is that the perturbation or the query point $x+\delta w$ may not remain in the feasible set $\mathbb{X}$. To avoid this, one can first transfer $x$ to an interior of $\mathbb{X}$ by a transformation of the form
\begin{align*}
x\mapsto x-\frac{\delta}{r}(x-p),
\end{align*}
where $p\in\mathbb{X}$ is an interior point of $\mathbb{X}$ and $r>\delta$ is selected such that the ball $\mathcal{B}_r(p)$ centered at $p$ with radius $r$ is entirely contained in $\mathbb{X}$. With this transformation, it can ensure that the query point
\begin{align*}
\hat{x}:&=x-\frac{\delta}{r}(x-p)+\delta w\\&=(1-\delta/r)x+(\delta/r)(p+rw)
\end{align*}
is in the convex set $\mathbb{X}$ since $p+rw\in\mathcal{B}_r(p)\subseteq\mathbb{X}$.

Based on the above introduced one-point gradient estimator, the process for estimating the payoff gradient of each player at time $t$ is given as follows:
\begin{itemize}
\item[1)] Each player $i\in[N]$ chooses a point $x_{i,t}\in\Omega_i$ and selects a perturbation direction $w_{i,t}$ from $\{\pm{\bf e}_1,\pm{\bf e}_2,\ldots,\pm{\bf e}_{n_i}\}$ uniformly at random. Subsequently, player $i$ adopts a strategy
    \begin{align}
    \hat{x}_{i,t}=x_{i,t}+\delta_{i,t}w_{i,t}+(\delta_{i,t}/r_{i})(p_{i}-x_{i,t}),
    \end{align}
    and then receives the corresponding reward or payoff value $\hat{J}_{i,t}:=J_{i,t}(\hat{x}_{1,t},\ldots,\hat{x}_{N,t})$ and the local constraint function value $\hat{g}_{ij,t}:=g_{ij,t}(\hat{x}_{i,t})$. Denote $\hat{g}_{i,t}:=col(\hat{g}_{i1,t},\ldots,\hat{g}_{im,t})$ and $\hat{g}_t:=\sum_{i=1}^N\hat{g}_{i,t}$.
\item[2)] Each player $i\in[N]$ makes an estimate of its payoff gradient and local constraint function gradient:
\begin{align}
\hat{\nabla}_iJ_{i,t}&:=\frac{n_i}{\delta_{i,t}}\hat{J}_{i,t}w_{i,t},\label{equ80}\\
\hat{\nabla}g_{ij,t}&:=\frac{n_i}{\delta_{i,t}}\hat{g}_{ij,t}w_{i,t}.\label{equ81}
\end{align}
Denote $\hat{\nabla}g_{i,t}:=(\hat{\nabla}g_{i1,t},\ldots,\hat{\nabla}g_{im,t})$.
\end{itemize}

Next, relying on Algorithm \ref{alg1}, a payoff-based learning algorithm is devised, see Algorithm \ref{alg2}.
\begin{algorithm}[!htbp]\caption{Payoff-based Online Primal-Dual Dynamic Mirror Descent}\label{alg2}
Each player $i$ maintains vector variables $x_{i,t}\in\mathbb{R}^{n_i}$ and $\lambda_{i,t}\in\mathbb{R}^{m}$ at iteration $t\in[T]$. Fix an interior point $p_{i}$ of $\Omega_i$, and choose $r_i>0$ such that $\mathcal{B}_{r_i}(p_i)\subseteq\Omega_i$. Let $\delta_{i,1}<r_i$.

{\bf Initialization:} For any $i\in[N]$, initialize $x_{i,1}\in\Omega_i$ arbitrarily and $\lambda_{i,1}={\bf0}_m$.

{\bf Iteration:} For $t\geq 1$, every player $i$ performs the following update:
\begin{align}
x_{i,t+1}&=\arg\min\limits_{x\in\Omega_i}\{\alpha_t\langle x,\hat{\nabla}_iJ_{i,t}+\hat{\nabla}g_{i,t}\tilde{\lambda}_{i,t}\rangle\nonumber\\
&~~~~~~~~~~+D_{\phi_i}(x,x_{i,t})\},\label{equ82}\\
\hat{x}_{i,t+1}&=x_{i,t+1}+\delta_{i,t}w_{i,t+1}+(\delta_{i,t+1}/r_{i})(p_{i}-x_{i,t+1}),\label{}\\
\lambda_{i,t+1}&=[\tilde{\lambda}_{i,t}+\gamma_t(\hat{g}_{i,t}-\beta_t\tilde{\lambda}_{i,t})]_+,\label{}
\end{align}
where $\tilde{\lambda}_{i,t}:=\sum_{j=1}^Na_{ij}\lambda_{j,t}$,
$a_{ij}$ is the $(i,j)$th element of $A$, and $\alpha_t,\beta_t,\gamma_t$, satisfying $\alpha_0=\beta_0=\gamma_0=1$, are the stepsizes to be determined.
\end{algorithm}

In Algorithm \ref{alg2}, each player $i\in[N]$ chooses the sampling radius $\delta_{i,t}$, $p_{i}$ and $r_{i}$ independently, only guaranteeing that $\mathcal{B}_{r_{i}}(p_{i})\subseteq\Omega_i$ and $r_{i}>\delta_{i,t}$, which indeed can be ensured by $r_i>\delta_{i,1}$, the particular selection of $r_i$, and the decrease of $\delta_{i,t}$. %Therefore, the initial value of $\delta_{i,t}$ should be chosen small enough by considering the radius of the biggest ball contained in the convex set $\Omega_i$, denoted as $o_i$.
For example, if the convex set $\Omega_i$ contains a ball $\mathcal{B}_h(0)$, then $p_i$ can be selected as the original point and $r_i$ can be chosen as $h$. $w_{i,t}$ is uniformly randomly chosen from $\{\pm{\bf e}_1,\ldots,\pm{\bf e}_N\}$. Let $\mathfrak{F}_t$ be the $\sigma$-algebra generated by $(w_{1,t},\ldots,w_{N,t})$ and $\mathcal{F}_t:=\bigcup_{s=1}^t\mathfrak{F}_s$.

\begin{lemma}[\cite{duvocelle2022multiagent}]\label{lemma4}
	The estimators in (\ref{equ80}) and (\ref{equ81}) satisfy
	\begin{align}
		\|\mathbf{E}[\hat{\nabla}_iJ_{i,t}-\nabla_iJ_{i,t}(x_t)]\|&=\mathcal{O}(\bar{\delta}_{t}^2/\delta_{i,t}),\\
		\|\mathbf{E}[\hat{\nabla}g_{i,t}-\nabla g_{i,t}(x_{i,t})]\|&=\mathcal{O}(\delta_{i,t}),\\
		\mathbf{E}[\|\hat{\nabla}_iJ_{i,t}\|^2]&=\mathcal{O}(1/\delta_{i,t}^2),\\
		\mathbf{E}[\|\hat{\nabla}g_{i,t}\|^2]&=\mathcal{O}(1/\delta_{i,t}^2),
	\end{align}
where $\bar{\delta}_{t}:=\max_{i\in[N]}\delta_{i,t}$.
\end{lemma}

In what follows, by leveraging the analysis on establishing the regret bounds in Section \ref{section3}, we derive the corresponding result on Algorithm \ref{alg2}.

\begin{theorem}\label{thm4}
If Assumptions \ref{assump1}--\ref{assump3} hold, then for each $i\in[N]$ and the sequence $\{x_{i,1},\ldots,x_{i,T}\}$ generated by Algorithm \ref{alg2} with
$$\alpha_t=\frac{1}{t^{d_1}},\beta_t=\frac{1}{t^{d_2}},\gamma_t=\frac{1}{t^{1-d_2}},\delta_{i,t}=\mathcal{O}(\frac{1}{t^{d_3}})$$ where $\alpha_0=\beta_0=\gamma_0=1$, $0<d_1<1$, $d_1>2d_2>0$ and $d_2<d_3<d_1$, there hold
\begin{align}
&\mathbf{E}[Reg_i(T)]\nonumber\\
&=\mathcal{O}(T^{\max\{d_1,1-d_1+d_3,1-d_1+2d_2,1+d_2-d_3\}}),\label{equ89}\\
&\mathbf{E}[R_g(T)]\nonumber\\
&=\mathcal{O}(T^{\max\{\frac{1}{2}+\frac{d_1}{2}-\frac{d_2}{2},1-\frac{d_1}{2}-\frac{d_2}{2}+\frac{d_3}{2},1-\frac{d_1}{2}+\frac{d_2}{2},1-\frac{d_3}{2}\}}).\label{equ90}
\end{align}
\end{theorem}

{\em Proof:}
See Appendix \ref{F}.
\hfill$\blacksquare$

\begin{corollary}
 Under Assumptions \ref{assump1}--\ref{assump3}, Algorithm \ref{alg2} achieves sublinearly bounded regrets and constraint violation, i.e., $Reg_i(T)={\bf o}(T)$ and $R_g(T)={\bf o}(T)$, if $0<2d_2<d_1<1$ and $d_1>d_3>d_2>0$.
\end{corollary}

\begin{corollary}
 Under Assumptions \ref{assump1}--\ref{assump3}, Algorithm \ref{alg2} can achieve the bounds on the regrets and constraint violation as
\begin{align*}
Reg_i(T)&=\mathcal{O}(T^{\frac{3}{4}}),\\
R_g(T)&=\mathcal{O}(T^{\frac{3}{4}}).
\end{align*}
\end{corollary}

{\em Proof:}
By letting $d_1=1-d_1+d_3=1-d_1+2d_2$, it is obtained that $d_1=1/2+d_2$ and $d_3=2d_2$. Then, the bound of $Reg_i(T)$ is derived as $Reg_i(T)=\mathcal{O}(T^{\max\{1/2+d_2,1-d_2\}})$. Making $1/2+d_2=1-d_2$ yields that $d_2=1/4$. Via a simple computation, one has
$Reg_i(T)=\mathcal{O}(T^{\frac{3}{4}})$ and $R_g(T)=\mathcal{O}(T^{\frac{3}{4}})$.
\hfill$\blacksquare$

\begin{remark}
In Algorithm \ref{alg2}, only payoff information and local constraint function values are used, which is usually easy to implement, but at the expense of a little bit worse bounds on the regrets and constraint violation compared with the full information case as shown in Corollary \ref{corollary2}. Moreover, in the one-point bandit feedback case, the convergence in expectation of the play sequence generated by Algorithm \ref{alg2} can be obtained similar to  that in Section \ref{section4}, whose details are omitted here.
\end{remark}

\section{A Numerical Example}\label{section6}
\begin{figure}[!ht]
 \centering
  \includegraphics[width=3.2in]{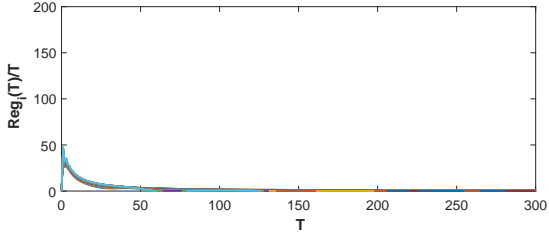}
  \caption{The trajectories of $Reg_i(T)/T$, $i\in[N]$ by Algorithm \ref{alg1}.}\label{fig1}
\end{figure}

\begin{figure}[!ht]
 \centering
  \includegraphics[width=3.2in]{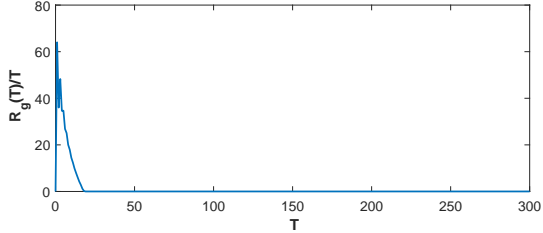}
  \caption{The trajectories of $R_g(T)/T$, $i\in[N]$ by Algorithm \ref{alg1}.}\label{fig2}
\end{figure}

In this section, a time-varying Nash-Cournot game $\Gamma(\mathcal{V},\Omega_t,J_t)$ with production constraints and market capacity constraints is considered to illustrate the feasibility of the obtained algorithm. Similar to \cite{lu2020online}, we consider a Nash-Cournot game, in which there are $N=20$ firms communicating with each other via a connected graph $\mathcal{G}$. Denote by $x_{i,t}\in\mathbb{R}$ the quality produced by firm $i$ at time $t$. In view of some uncertain and changeable factors such as marginal costs and demand for orders, the demand cost and the production cost may be time-varying. Assume that the production cost and the demand price of firm $i$ are $p_{i,t}(x_{i,t})=x_{i,t}(\sin(t/12)+1)$ and $d_{i,t}(x_{t})=22+i/9-0.5i\sin(t/12)-\sum_{j=1}^Nx_{j,t}$, respectively. Then, the overall cost function of firm $i$ is $J_{i,t}(x_{i,t},x_{-i,t})=p_{i,t}(x_{i,t})-x_{i,t}d_{i,t}(x_{t})$ for $i\in[N]$ and $t\in[T]$. In addition, the production quality constraint of firm $i$ is $x_{i,t}\in\Omega_i:=[0,30]$, while the market capacity constraint is modeled as a coupled inequality constraint $\sum_{i=1}^Nx_{i,t}\leq\sum_{i=1}^Nb_{i,t}$, where $b_{i,t}=2+sin(t/12)$ is the local bound available to firm $i$. In the offline and centralized setting, the GNE can be calculated as $x_{i,t}^*=P_{\Omega_i}(\xi_{i,t})$, where $\xi_{i,t}:=\frac{1}{9}(i-1)+(5-1/21-i/2)\sin\frac{t}{12}$. In the online and decentralized setting, for Algorithm \ref{alg1}, set initial states $x_{i,0}\in\Omega_{i}$ randomly, $\lambda_{i,1}={\bf0}_m$, and choose $a_1=0.8$ and $a_2=0.3$. $Reg_i(T)/T$, $i\in[N]$ and $R_g(T)/T$ are shown in Fig. \ref{fig1} and Fig. \ref{fig2}, respectively.
%Similarly, by Algorithm \ref{alg2}, set initial states $x_{i,0}\in\Omega_{i}$ randomly, $x_{ih,1}={\bf0}_{n_h}(h\neq i)$, and $\lambda_{i,1}={\bf0}_m$. Choose $a_1=0.2$ and $a_2=0.8$. Then, the expected regrets ${\bf E}[Reg_i(T)]/T$, $i\in[N]$, and the expected violation ${\bf E}[R_g(T)]/T$ are shown in Fig. \ref{fig3} and Fig. \ref{fig4}, respectively. The average is taken over 100 realizations.
From these figures, one can see that the average regret $Reg_i(T)/T$ and the average violation $R_g(T)/T$ decay to zero as iteration goes on. That is, the regrets $Reg_i(T)$, $i\in[N]$, and the violation $R_g(T)$ increase sublinearly, which are consistent with theoretical results in Section \ref{section3}.
%On the other hand, it can be seen that the convergence rates for the regrets and the constraint violation generated by Algorithm \ref{alg2} are slower than those generated by Algorithm \ref{alg1}, which is reasonable as one-point gradient estimator utilizes less information of the considered game.

\begin{figure}[!ht]
 \centering
  \includegraphics[width=3.2in]{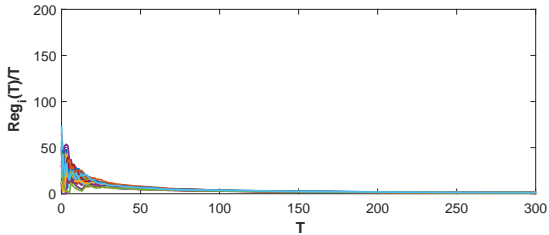}
  \caption{The trajectory of $Reg_i(T)/T$, $i\in[N]$ by Algorithm \ref{alg2}.}\label{fig4}
\end{figure}

\begin{figure}[!ht]
 \centering
  \includegraphics[width=3.2in]{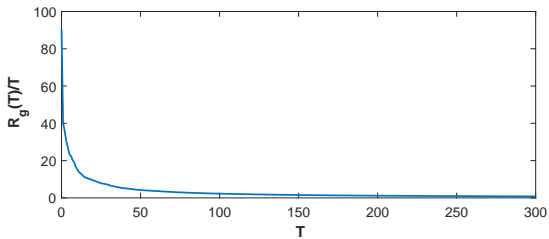}
  \caption{The trajectories of $R_g(T)/T$, $i\in[N]$ by Algorithm \ref{alg2}.}\label{fig5}
\end{figure}

When players can only receive the rewards or payoff rather than the information on gradients of payoff and constraint functions after decisions are made, by the payoff-based learning algorithm (Algorithm \ref{alg2}), the simulations on $Reg_i(T)/T$ and $R_g(T)$ are shown in Fig. \ref{fig4} and Fig. \ref{fig5}, respectively. Here, the interior point $p_{i}$ of $\Omega_i$ is selected as $3$, the radius $r_i$ is $1.5$ and the query radius $\delta_t$ is $\mathcal{O}(1/t^{0.4})$. From the simulations, one can see that the regrets and constraint violation increase sublinearly, which is consistent with theoretical results in Section \ref{section5}. Moreover, one also can see that the bounds of regrets and constraint violation are a little larger than those of the case of gradient accuracy feedback.

\begin{figure}[!ht]
 \centering
  \includegraphics[width=3.2in]{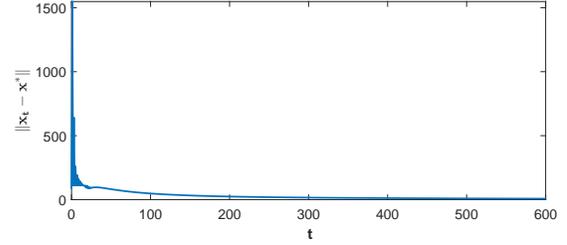}
  \caption{The convergence of $\|x_t-x^*\|$ by Algorithm \ref{alg1}.}\label{fig3}
\end{figure}

\begin{figure}[!ht]
 \centering
  \includegraphics[width=3.2in]{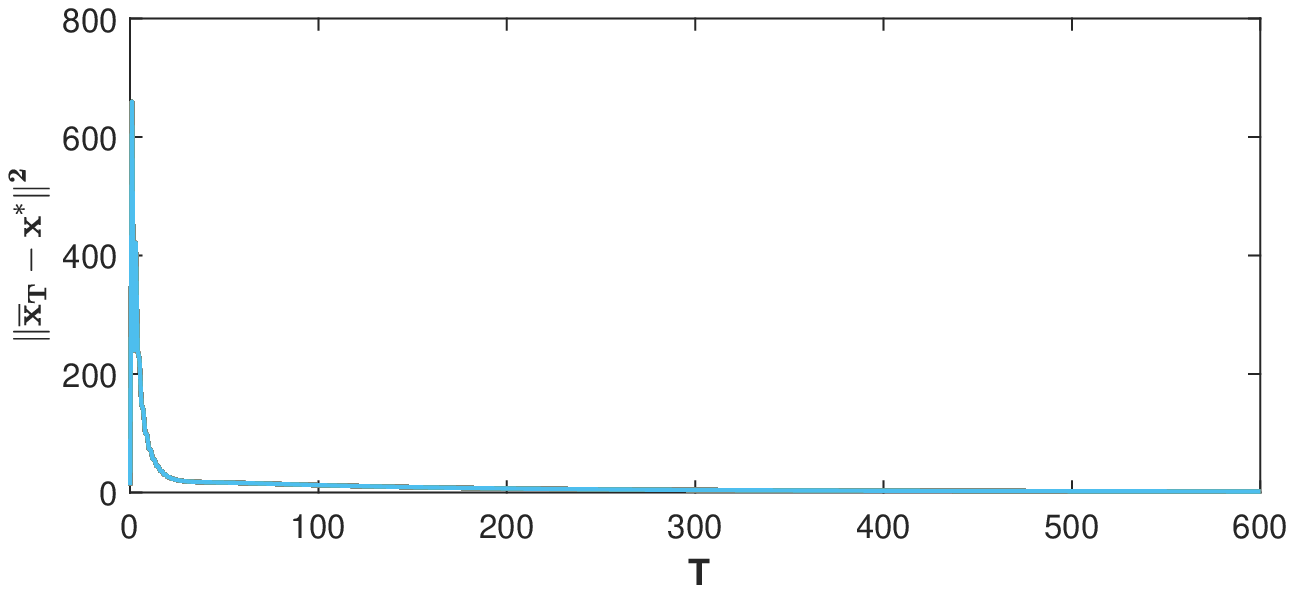}
  \caption{The convergence of $\|\overline{x}_T-x^*\|^2$ by Algorithm \ref{alg1}.}\label{fig33}
\end{figure}

If the term $\sin(t/12)$ is replaced by $\sin(12/t)$, then online game $\Gamma_t$ converges to a strongly monotone game $\Gamma$ with payoff function $J_{i}(x_{i},x_{-i})=x_i-x_i(22+i/9-\sum_{j=1}^Nx_j)$ for $i\in[N]$ and the coupled constraint $\sum_{i=1}^Nx_{i,t}\leq2N$. In this case, the errors $\|x_t-x^*\|$ and $\|\sum_{t=1}^Tx_t/T-x^*\|^2$ are shown in Fig. \ref{fig3} and Fig. \ref{fig33}, respectively, where $x^*$ is the unique variational GNE of $\Gamma$. These simulations demonstrate the theoretical results in Section \ref{section4}.
\section{Conclusion}\label{section7}
In this paper, decentralized learning for online game with time-varying cost functions and time-varying general convex constraints was investigated. A novel decentralized online algorithm was devised based on the mirror descent and primal-dual strategy. It was rigorously proved that the proposed algorithm could achieve sublinearly bounded regrets and constraint violation by appropriately choosing decaying stepsizes. The convergence property of the no-regret learning was also analyzed. Furthermore, the one-point bandit feedback case was also studied. And all theoretical results have been validated by a numerical example.
%Future research of interest is to develop new online algorithms to improve the bounds of dynamic regrets and constraint violation, and study the regret defined in \cite{hsieh2021adaptive} for online games considered in this paper.

\section*{Acknowledgment}
%The authors are grateful to the Editor, the Associate Editor, and the anonymous reviewers for their insightful suggestions, which certainly improves the quality of this work.
The authors are grateful to Prof. Kaihong Lu for helpful discussion.

\begin{appendix}
\subsection{Proof of Lemma \ref{lemma1}}\label{A}
It can be easily proved that (\ref{equ18}) and (\ref{equ19}) hold by mathematical induction \cite{yi2020distributed}. Therefore, it suffices to prove (\ref{equ20}) and (\ref{equ21}). From the iteration (\ref{equ17}), one has
\begin{align}
	\lambda_{i,t+1}=\sum_{j=1}^Na_{ij}\lambda_{j,t}+\epsilon_{i,t},
\end{align}
where $\epsilon_{i,t}:=[\tilde{\lambda}_{i,t}+\gamma_t(C_{i,t}-\beta_t\tilde{\lambda}_{i,t})]_+-\tilde{\lambda}_{i,t}$, $i\in[N]$. Denote $\lambda_t:=col(\lambda_{1,t},\ldots,\lambda_{N,t})$ and $\epsilon_t:=col(\epsilon_{1,t},\ldots,\epsilon_{N,t})$. Then
\begin{align}
	\lambda_{t+1}-{\bf1}_N\otimes\overline{\lambda}_{t+1}&=((A-\frac{1}{N}{\bf1}_N{\bf1}_N^{\top})\otimes I_m)(\lambda_{t}-{\bf1}_N\otimes\overline{\lambda}_t)\nonumber\\
	&~~~+((I-\frac{1}{N}{\bf1}_N{\bf1}_N^{\top})\otimes I_m)\epsilon_t. \label{equ23}
\end{align}
Under Assumption \ref{assump1}, one has $0<\sigma<1$, where $\sigma=\|A-\frac{1}{N}{\bf1}_N{\bf1}_N^{\top}\|$. Consequently, taking norm on both sides of (\ref{equ23}) yields
\begin{align}
	\|\lambda_{t+1}-{\bf1}_N\otimes\overline{\lambda}_{t+1}\|
	&\leq\sigma\|\lambda_{t}-{\bf1}_N\otimes\overline{\lambda}_t\|+\|\epsilon_t\|.
\end{align}
Note that
\begin{align}
	\|\epsilon_{i,t}\|&= \Big\|\Big[\tilde{\lambda}_{i,t}+\gamma_t(C_{i,t}-\beta_t\tilde{\lambda}_{i,t})\Big]_+-\tilde{\lambda}_{i,t}\Big\|\nonumber\\
	&\leq \gamma_t\|C_{i,t}-\beta_t\tilde{\lambda}_{i,t}\|\nonumber\\
	&\leq \gamma_t\|C_{i,t}\|+\gamma_t\beta_t\|\tilde{\lambda}_{i,t}\|\nonumber\\
	&\leq 2L\gamma_t,\label{}
\end{align}
where the first inequality is derived based on $\|[a]_+-[b]_+\|\leq\|a-b\|$ for any two vectors $a,b$ with the same dimension, and the third inequality is obtained based on (\ref{equ2}) in Assumption \ref{assump2} and (\ref{equ19}). Then,
\begin{align*}
	\|\lambda_{t+1}-{\bf1}_N\otimes\overline{\lambda}_{t+1}\|\leq\sigma\|\lambda_{t}-{\bf1}_N\otimes\overline{\lambda}_t\|+2\sqrt{N}L\gamma_t.
\end{align*}
Combined with $\lambda_{1}-{\bf1}_N\otimes\overline{\lambda}_{1}={\bf0}_{Nm}$, (\ref{equ20}) is thus proved.

For (\ref{equ21}), it can be obtained from (\ref{equ17}) that for any $\lambda\in\mathbb{R}^m_+$,
\begin{align}
	\|\lambda_{i,t+1}-\lambda\|^2
	&=\left\|\left[\tilde{\lambda}_{i,t}+\gamma_t(C_{i,t}-\beta_t\tilde{\lambda}_{i,t})\right]_+-\lambda\right\|^2\nonumber\\
	&\leq\left\|\tilde{\lambda}_{i,t}+\gamma_t(C_{i,t}-\beta_t\tilde{\lambda}_{i,t})-\lambda\right\|^2\nonumber\\
	&=\|\tilde{\lambda}_{i,t}-\lambda\|^2+\gamma_t^2\|C_{i,t}-\beta_t\tilde{\lambda}_{i,t}\|^2\nonumber\\
	&~~~+2\gamma_t(\tilde{\lambda}_{i,t}-\lambda)^{\top}(C_{i,t}-\beta_t\tilde{\lambda}_{i,t})\nonumber\\
	&\leq\|\tilde{\lambda}_{i,t}-\lambda\|^2+4L^2\gamma_t^2\nonumber\\
	&~~~+2\gamma_t(\tilde{\lambda}_{i,t}-\lambda)^{\top}(C_{i,t}-\beta_t\tilde{\lambda}_{i,t}),\label{equ27}
\end{align}
where the last inequality is based on (\ref{equ2}) in Assumption \ref{assump2} and (\ref{equ19}).

For the last term in (\ref{equ27}),
\begin{align}
	&2\gamma_t(\tilde{\lambda}_{i,t}-\lambda)^{\top}(C_{i,t}-\beta_t\tilde{\lambda}_{i,t})\nonumber\\
	&=2\gamma_t(\tilde{\lambda}_{i,t}-\overline{\lambda}_t)^{\top}C_{i,t}+2\gamma_t(\overline{\lambda}_t-\lambda)^{\top}C_{i,t}\nonumber\\
	&~~~-2\beta_t\gamma_t(\tilde{\lambda}_{i,t}-\lambda)^{\top}\tilde{\lambda}_{i,t}\nonumber\\
	&\leq2\gamma_t\|\tilde{\lambda}_{i,t}-\overline{\lambda}_t\|\|C_{i,t}\|+2\gamma_t(\overline{\lambda}_t-\lambda)^{\top}C_{i,t}\nonumber\\
	&~~~+\beta_t\gamma_t(\|\lambda\|^2-\|\tilde{\lambda}_{i,t}-\lambda\|^2)\nonumber\\
	&\leq4\sqrt{N}L^2\gamma_t\sum\limits_{s=0}^{t-1}\sigma^{s}\gamma_{t-1-s}+2\gamma_t(\overline{\lambda}_t-\lambda)^{\top}C_{i,t}\nonumber\\
	&~~~+\beta_t\gamma_t(\|\lambda\|^2-\|\tilde{\lambda}_{i,t}-\lambda\|^2). \label{equ28}
\end{align}
Substituting (\ref{equ28}) into (\ref{equ27}) and summing over $t=1,2,\ldots,T$, one can obtain (\ref{equ21}). \hfill$\blacksquare$

\subsection{Proof of Lemma \ref{lemma2}}\label{B}
For any $i\in[N]$, based on the optimality of ${x}_{i,t+1}$ in (\ref{equ16}), one can obtain that for any $x_i\in\Omega_i$,
\begin{align}
	&\left\langle{x}_{i,t+1}-x_i,\alpha_t(V_{i,t}+G_{i,t}\tilde{\lambda}_{i,t})\right\rangle\nonumber\\
	&+\left\langle{x}_{i,t+1}-x_i,\nabla\phi_i({x}_{i,t+1})-\nabla\phi_i({x}_{i,t})\right\rangle\leq 0,\label{equ30}
\end{align}
where $\frac{\partial D_{\phi_i}(x,x_{i,t})}{\partial x}=\nabla\phi_i({x})-\nabla\phi_i({x}_{i,t})$ for $x\in\Omega_i$ is used. Then, it can be derived from (\ref{equ30}) that
\begin{align}
	&\alpha_t\left\langle{x}_{i,t+1}-{x}_{i},V_{i,t}+G_{i,t}\tilde{\lambda}_{i,t}\right\rangle\nonumber\\
	&\leq \left\langle{x}_{i}-{x}_{i,t+1},\nabla\phi_i({x}_{i,t+1})-\nabla\phi_i({x}_{i,t})\right\rangle\nonumber\\
	&=D_{\phi_i}({x}_{i},x_{i,t})-D_{\phi_i}({x}_{i},{x}_{i,t+1})-D_{\phi_i}({x}_{i,t+1},x_{i,t})\nonumber\\
	&\leq D_{\phi_i}({x}_{i},x_{i,t})-D_{\phi_i}({x}_{i},{x}_{i,t+1})\nonumber\\
	&~~~-\frac{\mu_0}{2}\|{x}_{i,t+1}-x_{i,t}\|^2,\label{equ31}
\end{align}
where the equality is derived based on (\ref{equ11}) and the last inequality is based on (\ref{equ10}).
Then, rearranging (\ref{equ31}) yields
\begin{align}
	\alpha_t\left\langle{x}_{i,t}-{x}_{i},V_{i,t}\right\rangle
	&\leq\alpha_t\left\langle{x}_{i,t}-x_{i,t+1},V_{i,t}\right\rangle\nonumber\\
	&~~~+\alpha_t\langle{x}_{i}-x_{i,t+1},G_{i,t}\tilde{\lambda}_{i,t}\rangle\nonumber\\
	&~~~+D_{\phi_i}({x}_{i},x_{i,t})-D_{\phi_i}({x}_{i},{x}_{i,t+1})\nonumber\\
	&~~~-\frac{\mu_0}{2}\|{x}_{i,t+1}-x_{i,t}\|^2
	.\label{equ32}
\end{align}
As to the first term on the right hand side of (\ref{equ32}), we have
\begin{align}
	\alpha_t\left\langle{x}_{i,t}-x_{i,t+1},V_{i,t}\right\rangle\leq\frac{M^2\alpha_t^2}{\mu_0}+\frac{\mu_0}{4}\|x_{i,t}-{x}_{i,t+1}\|^2,\label{equ33}
\end{align}
where the inequality is obtained by the Cauchy-Schwarz inequality and (\ref{equ3}) in Assumption \ref{assump2}.

For the second term on the right hand side of (\ref{equ32}), one has
\begin{align}
	&\alpha_t\left\langle{x}_{i}-{x}_{i,t+1},G_{i,t}\tilde{\lambda}_{i,t}\right\rangle\nonumber\\
	&=\alpha_t({x}_{i}-x_{i,t})^{\top}\nabla{g_{i,t}}(x_{i,t})\tilde{\lambda}_{i,t}\nonumber\\
	&~~~+\alpha_t({x}_{i,t}-x_{i,t+1})^{\top}\nabla{g_{i,t}}(x_{i,t})\tilde{\lambda}_{i,t}\nonumber\\
	&\leq\alpha_t\tilde{\lambda}_{i,t}^{\top}(g_{i,t}({x}_{i})-g_{i,t}(x_{i,t}))\nonumber\\
	&~~~+\alpha_t({x}_{i,t}-x_{i,t+1})^{\top}\nabla{g_{i,t}}(x_{i,t})\tilde{\lambda}_{i,t}\nonumber\\
	&=\alpha_t(\tilde{\lambda}_{i,t}-\overline{\lambda}_t)^{\top}(g_{i,t}(x_{i})-g_{i,t}(x_{i,t}))\nonumber\\
	&~~~+\alpha_t\overline{\lambda}_t^{\top}(g_{i,t}(x_{i})-g_{i,t}(x_{i,t}))\nonumber\\
	&~~~+\alpha_t({x}_{i,t}-x_{i,t+1})^{\top}\nabla{g_{i,t}}(x_{i,t})\tilde{\lambda}_{i,t}\nonumber\\
	&\leq2L\alpha_t\|\tilde{\lambda}_{i,t}-\overline{\lambda}_{t}\|+\alpha_t\overline{\lambda}_{t}^{\top}(g_{i,t}(x_{i})-g_{i,t}(x_{i,t}))\nonumber\\
	&~~~+\frac{\alpha_t^2\|\tilde{\lambda}_{i,t}\|^2\|\nabla{g_{i,t}}(x_{i,t})\|^2}{\mu_0}+\frac{\mu_0}{4}\|x_{i,t}-{x}_{i,t+1}\|^2\nonumber\\
	&\leq4\sqrt{N}L^2\alpha_t\sum\limits_{s=0}^{t-1}\sigma^{s}\gamma_{t-1-s}+\alpha_t\overline{\lambda}_{t}^{\top}(g_{i,t}(x_{i})-g_{i,t}(x_{i,t}))\nonumber\\
	&~~~+\frac{L^2M^2}{\mu_0}\frac{\alpha_t^2}{\beta_t^2}+\frac{\mu_0}{4}\|x_{i,t}-{x}_{i,t+1}\|^2,\label{equ34}
\end{align}
where the first inequality is derived relying on the convexity of $g_{ij,t}$ and $\tilde{\lambda}_{i,t}\geq0$, the second inequality is obtained by (\ref{equ2}) in Assumption \ref{assump2}, and the last inequality applies (\ref{equ20}) in Lemma \ref{lemma1} and (\ref{equ2}) in Assumption \ref{assump2}.

Substituting (\ref{equ33}) and (\ref{equ34}) into (\ref{equ32}) yields (\ref{equ29}). \hfill$\blacksquare$
\subsection{Proof of Lemma \ref{lemma3}}\label{C}
For (\ref{equ29}) in Lemma \ref{lemma2}, setting $x_i=\tilde{x}_i$, where $\tilde{x}_i$ is defined below (\ref{equ7}), one has
\begin{align}
	\langle x_{i,t}-\tilde{x}_i,V_{i,t}\rangle&\leq\frac{M^2}{\mu_0}\alpha_t+4\sqrt{N}L^2\sum_{s=0}^{t-1}\sigma^s\gamma_{t-1-s}\nonumber\\
	&+\frac{M^2L^2}{\mu_0}\frac{\alpha_t}{\beta_t^2}+\overline{\lambda}_t^{\top}(g_{i,t}(\tilde{x}_i)-g_{i,t}(x_{i,t}))\nonumber\\
	&+\frac{1}{\alpha_t}[D_{\phi_i}(\tilde{x}_i,x_{i,t})-D_{\phi_i}(\tilde{x}_i,x_{i,t+1})].\label{equ37}
\end{align}

Since $\tilde{x}_i\in\Omega_{i,t}(x_{-i,t})$, i.e., $g_{i,t}(\tilde{x}_i)+\sum_{j\neq i}g_{j,t}(x_{j,t})$, it is obtained that
\begin{align}
	&\overline{\lambda}_t^{\top}(g_{i,t}(\tilde{x}_i)-g_{i,t}(x_{i,t}))\nonumber\\
	&=\overline{\lambda}_t^{\top}(g_{i,t}(\tilde{x}_i)+\sum_{j\neq i}g_{j,t}(x_{j,t})-g_{t}(x_{t}))\nonumber\\
	&\leq-\overline{\lambda}_t^{\top}g_{t}(x_{t}).\label{equ38}
\end{align}
Substituting (\ref{equ38}) into (\ref{equ37}) yields
\begin{align}
	\langle x_{i,t}-\tilde{x}_i,V_{i,t}\rangle&\leq\frac{M^2}{\mu_0}\alpha_t+4\sqrt{N}L^2\sum_{s=0}^{t-1}\sigma^s\gamma_{t-1-s}\nonumber\\
	&~~~+\frac{1}{\alpha_t}[D_{\phi_i}(\tilde{x}_i,x_{i,t})-D_{\phi_i}(\tilde{x}_i,x_{i,t+1})]\nonumber\\
	&~~~+\frac{M^2L^2}{\mu_0}\frac{\alpha_t}{\beta_t^2}-\overline{\lambda}_t^{\top}g_{t}(x_{t}).\label{equ39}
\end{align}

On the other hand, by $\Xi_{t+1}$ defined in Lemma \ref{lemma1}, one obtains
\begin{align}
	-\frac{\Xi_{t+1}}{2\gamma_t}&=-\frac{1}{2}\sum\limits_{i=1}^N\left[\frac{1}{\gamma_t}\|\lambda_{i,t+1}-\lambda\|^2-\frac{1}{\gamma_{t-1}}\|\lambda_{i,t}-\lambda\|^2\right]\nonumber\\
	&~~~+\frac{1}{2}\left(\frac{1}{\gamma_t}-\frac{1}{\gamma_{t-1}}-\beta_t\right)\sum_{i=1}^N\|\lambda_{i,t}-\lambda\|^2. \label{equ40}
\end{align}
Summing over $t\in[T]$ gives that
\begin{align}
	&-\sum_{t=1}^T\frac{\Xi_{t+1}}{2\gamma_t}\nonumber\\
	&=-\frac{1}{2}\sum\limits_{i=1}^N\left[\frac{1}{\gamma_T}\|\lambda_{i,T+1}-\lambda\|^2-\frac{1}{\gamma_{0}}\|\lambda_{i,1}-\lambda\|^2\right]\nonumber\\
	&~~~+\frac{1}{2}\sum\limits_{t=1}^T\left(\frac{1}{\gamma_t}-\frac{1}{\gamma_{t-1}}-\beta_t\right)\sum_{i=1}^N\|\lambda_{i,t}-\lambda\|^2\nonumber\\
	&\leq\frac{N}{2}\|\lambda\|^2+\frac{1}{2}\sum\limits_{t=1}^T\left(\frac{1}{\gamma_t}-\frac{1}{\gamma_{t-1}}-\beta_t\right)\sum_{i=1}^N\|\lambda_{i,t}-\lambda\|^2,\label{equ41}
\end{align}
where the inequality is derived based on $\lambda_{i,1}={\bf0}_m$.

By (\ref{equ21}) in Lemma \ref{lemma1}, summing over $t\in[T]$ on both sides of (\ref{equ39}) and (\ref{equ21}), combining with (\ref{equ41}), one has
\begin{align}
	&\sum_{t=1}^T\langle x_{i,t}-\tilde{x}_i,V_{i,t}\rangle\nonumber\\
	&\leq\sum\limits_{t=1}^T\frac{1}{\alpha_t}\left[D_{\phi_i}(\tilde{x}_i,x_{i,t})-D_{\phi_i}(\tilde{x}_i,x_{i,t+1})\right]\nonumber\\
	&~~~+2(N+2)\sqrt{N}L^2\sum_{t=1}^T\sum_{s=0}^{t-1}\sigma^s\gamma_{t-1-s}+\frac{M^2}{\mu_0}\sum_{t=1}^T\alpha_t\nonumber\\
	&~~~+2NL^2\sum_{t=1}^T\gamma_t+\frac{L^2M^2}{\mu_0}\sum_{t=1}^T\frac{\alpha_t}{\beta_t^2}-\lambda^{\top}\sum_{t=1}^Tg_t(x_t)\nonumber\\
	&~~~+\frac{N}{2}(1+\sum_{t=1}^T\beta_t)\|\lambda\|^2\nonumber\\
	&~~~+\frac{1}{2}\sum\limits_{t=1}^T\left(\frac{1}{\gamma_t}-\frac{1}{\gamma_{t-1}}-\beta_t\right)\sum_{i=1}^N\|\lambda_{i,t}-\lambda\|^2. \label{equ42}
\end{align}

For the first term on the right hand side of (\ref{equ42}), one has
\begin{align}
	&\sum_{t=1}^T\frac{1}{\alpha_t}\left[D_{\phi_i}(\tilde{x}_i,x_{i,t})-D_{\phi_i}(\tilde{x}_i,x_{i,t+1})\right]\nonumber\\
	&=\sum_{t=1}^T\left[\frac{1}{\alpha_t}D_{\phi_i}(\tilde{x}_i,x_{i,t})-\frac{1}{\alpha_{t+1}}D_{\phi_i}(\tilde{x}_i,x_{i,t+1})\right]\nonumber\\
	&~~~+\sum_{t=1}^T\left[\frac{1}{\alpha_{t+1}}-\frac{1}{\alpha_t}\right]D_{\phi_i}(\tilde{x}_i,x_{i,t+1})\nonumber\\
	&\leq\frac{1}{\alpha_1}D_{\phi_i}(\tilde{x}_i,x_{i,1})+\sum_{t=1}^T\left[\frac{1}{\alpha_{t+1}}-\frac{1}{\alpha_t}\right]2LK\nonumber\\
	&\leq\frac{2LK}{\alpha_{T+1}},\label{equ43}
\end{align}
where Assumption \ref{assump3} and (\ref{equ2}) have been used to get the first inequality.

Note that
\begin{align}
	\sum_{t=1}^T\sum\limits_{s=0}^{t-1}\sigma^s\gamma_{t-s-1}
	&=\sum_{t=1}^T\gamma_{t-1}\sum\limits_{s=0}^{T-t}\sigma^s\nonumber\\
	&\leq\frac{1}{1-\sigma}\sum_{t=1}^T\gamma_{t-1},\label{equ44}\\
	Reg_i(T)&\leq\sum_{t=1}^T\langle x_{i,t}-\tilde{x}_i,V_{i,t}\rangle,
\end{align}
then (\ref{equ35}) can be derived by making $\lambda={\bf0}_m$ in (\ref{equ42}).

Let $\lambda=\lambda_c=\frac{2\left[\sum_{t=1}^Tg_t(x_t)\right]_+}{B_3(T)}$, then one has from  (\ref{equ42}), (\ref{equ43}) and (\ref{equ44}) that
\begin{align}
	&\sum_{t=1}^T\langle x_{i,t}-\tilde{x}_i,V_{i,t}\rangle\leq B_1(T)+B_4(T)-\frac{1}{B_3(T)}R_g^2,
\end{align}
from which, it can be derived that
\begin{align}
	R_g^2(T)\leq B_1(T)B_3(T)+B_3(T)B_4(T),
\end{align}
where $\langle x_{i,t}-\hat{x}_{i},V_{i,t}\rangle\geq0$ is used.
Lemma \ref{lemma3} is proved. \hfill$\blacksquare$

\subsection{Proof of Theorem \ref{thm2}}\label{D}
This theorem is proved by splitting two steps: first prove that $\sum_{i=1}^ND_{\phi_i}(x_i^*,x_{i,t})$ converges to some finite value; and then prove that $x_t$ converges to $x^*$ by the mathematical induction method.

First, for any $i\in[N]$, let $x_i=x_i^*$ in (\ref{equ29}) of Lemma \ref{lemma2}. One has
\begin{align}
	\langle &x_{i,t}-x_i^*,V_{i,t}\rangle\leq\frac{M^2}{\mu_0}\alpha_t+4\sqrt{N}L^2\sum_{s=0}^{t-1}\sigma^s\gamma_{t-1-s}\nonumber\\
	&~~~~~~~~~~~~~~~~+\frac{M^2L^2}{\mu_0}\frac{\alpha_t}{\beta_t^2}+\overline{\lambda}_t^{\top}(g_{i,t}(x_i^*)-g_{i,t}(x_{i,t}))\nonumber\\
	&~~~~~~~~~~~~~~~~+\frac{1}{\alpha_t}[D_{\phi_i}(x_i^*,x_{i,t})-D_{\phi_i}(x_i^*,x_{i,t+1})],
\end{align}
that is,
\begin{align}
	D_{\phi_i}(x_i^*,x_{i,t+1})&\leq D_{\phi_i}(x_i^*,x_{i,t})+\alpha_t\langle x_i^*-x_{i,t},V_{i,t}\rangle\nonumber\\
	&~~~+\frac{M^2}{\mu_0}\alpha_t^2+4\sqrt{N}L^2\alpha_t\sum_{s=0}^{t-1}\sigma^s\gamma_{t-1-s}\nonumber\\
	&~~~+\alpha_t\overline{\lambda}_t^{\top}(g_{i,t}(x_i^*)-g_{i,t}(x_{i,t}))\nonumber\\
	&~~~+\frac{M^2L^2}{\mu_0}\frac{\alpha_t^2}{\beta_t^2}.
\end{align}
Summing over $i\in[N]$ yields
\begin{align}
	&\sum_{i=1}^ND_{\phi_i}(x_i^*,x_{i,t+1})\nonumber\\
	&\leq\sum_{i=1}^ND_{\phi_i}(x_i^*,x_{i,t})+\alpha_t\sum_{i=1}^N\langle x_i^*-x_{i,t},V_{i,t}\rangle\nonumber\\
	&~~~+\frac{NM^2}{\mu_0}\alpha_t^2+4N\sqrt{N}L^2\alpha_t\sum_{s=0}^{t-1}\sigma^s\gamma_{t-1-s}\nonumber\\
	&~~~+\alpha_t\overline{\lambda}_t^{\top}(g_{t}(x^*)-g_{t}(x_{t}))+\frac{NM^2L^2}{\mu_0}\frac{\alpha_t^2}{\beta_t^2}.\label{equ58}
\end{align}
Note that
\begin{align}
	&\sum_{i=1}^N\langle x_i^*-x_{i,t},V_{i,t}\rangle\nonumber\\
	&=\langle x^*-x_{t},F(x_t)-F(x^*)\rangle+\langle x^*-x_t,F_t(x_t)-F(x_t)\rangle\nonumber\\
	&~~~+\langle x^*-x_t,F(x^*)\rangle\nonumber\\
	&\leq-\mu\|x_t-x^*\|^2+\langle x^*-x_t,F_t(x_t)-F(x_t)\rangle\nonumber\\
	&~~~+\langle x^*-x_t,F(x^*)+\nabla g(x^*)\lambda^*\rangle+\langle x_t-x^*,\nabla g(x^*)\lambda^*\rangle\nonumber\\
	&\leq-\mu\|x_t-x^*\|^2+\|x^*-x_t\|\|F_t(x_t)-F(x_t)\|\nonumber\\
	&~~~+(\lambda^*)^{\top}(g(x_t)-g(x^*))\nonumber\\
	%&\leq-\mu\|x_t-x^*\|^2+2L\sum_{i=1}^NH_{i,t}+(\lambda^*)^{\top}g(x_t)\nonumber\\
	&\leq-\mu\|x_t-x^*\|^2+2L\sum_{i=1}^NH_{i,t}+(\lambda^*)^{\top}(g(x_t)-g_t(x_t))\nonumber\\
	&~~~+(\lambda^*)^{\top}g_t(x_t)\nonumber\\
	&\leq-\mu\|x_t-x^*\|^2+2L\sum_{i=1}^NH_{i,t}+\Lambda K_t\nonumber\\
	&~~~+(\lambda^*)^{\top}g_t(x_t),\label{equ59}
\end{align}
where the first inequality is obtained based on the $\mu$-strong monotonicity of $F$, the second inequality is derived by the fact that $(x_i^*,\lambda^*)$ is a saddle point of the Lagrangian function $\mathcal{L}_i(x_i,\lambda;x_{-i})$,  the convexity of $g_{ij}$ and nonnegativity of $\lambda^*$, the third inequality is got following the KKT condition $\langle\lambda^*,g(x^*)\rangle=0$ and (\ref{equ2}) in Assumption \ref{assump2}, and the last inequality is deduced by (\ref{equ54}).

In addition, by $g(x^*)\leq0$ and $\overline{\lambda}_t\geq0$, one has
\begin{align}
	\overline{\lambda}_t^{\top}g_{t}(x^*)
	&=\overline{\lambda}_t^{\top}(g_{t}(x^*)-g(x^*))+\overline{\lambda}_t^{\top}g(x^*)\nonumber\\
	&\leq\overline{\lambda}_t^{\top}(g_{t}(x^*)-g(x^*))\nonumber\\
	&\leq\frac{L}{\beta_t}K_t,\label{equ60}
\end{align}
where Lemma \ref{lemma1} and the notation $K_t$ have been used to derive the last inequality.

Substituting (\ref{equ59}) and (\ref{equ60}) into (\ref{equ58}) yields
\begin{align}
	&\sum_{i=1}^ND_{\phi_i}(x_i^*,x_{i,t+1})\nonumber\\
	&\leq\sum_{i=1}^ND_{\phi_i}(x_i^*,x_{i,t})-\mu\alpha_t\|x_t-x^*\|^2+2L\alpha_t\sum_{i=1}^NH_{i,t}\nonumber\\
	&~~~+\Lambda\alpha_t K_t+\alpha_t(\lambda^*-\overline{\lambda}_t)^{\top}g_t(x_t)+\frac{NM^2}{\mu_0}\alpha_t^2+2\alpha_tc(t)\nonumber\\
	&~~~+\frac{NM^2L^2}{\mu_0}\frac{\alpha_t^2}{\beta_t^2}+\frac{L\alpha_t}{\beta_t}K_t.\label{equ61}
\end{align}

Combining with (\ref{equ21}) in Lemma \ref{lemma1}, by making $\lambda=\lambda^*$, one has from (\ref{equ61}) that
\begin{align}
	\sum_{i=1}^ND_{\phi_i}(x_i^*,x_{i,t+1})&\leq\sum_{i=1}^ND_{\phi_i}(x_i^*,x_{i,t})-\mu\alpha_t\|x_t-x^*\|^2\nonumber\\
	&~~~+\mathcal{E}_t,\label{equ62}
\end{align}
where $\mathcal{E}_t:=2L\alpha_t\sum_{i=1}^NH_{i,t}+\Lambda\alpha_t K_t+\frac{NM^2}{\mu_0}\alpha_t^2+3\alpha_tc(t)+\frac{NM^2L^2}{\mu_0}\frac{\alpha_t^2}{\beta_t^2}+\frac{L\alpha_t}{\beta_t}K_t+2NL^2\alpha_t\gamma_t+\frac{N\Lambda}{2}\alpha_t\beta_t-\frac{\Xi_{t+1}\alpha_t}{2\gamma_t}$.

Similar to (\ref{equ40}) and (\ref{equ41}), one can derive that
\begin{align*}
	&-\sum_{t=1}^T\frac{\Xi_{t+1}\alpha_t}{2\gamma_t}\nonumber\\
	&\leq-\frac{1}{2}\sum_{t=1}^T\sum\limits_{i=1}^N\left[\frac{\alpha_t}{\gamma_t}\|\lambda_{i,t+1}-\lambda^*\|^2-\frac{\alpha_{t-1}}{\gamma_{t-1}}\|\lambda_{i,t}-\lambda^*\|^2\right]\nonumber\\
	&~~~+\frac{1}{2}\sum\limits_{t=1}^T\alpha_t\left(\frac{1}{\gamma_t}-\frac{1}{\gamma_{t-1}}-\beta_t\right)\sum_{i=1}^N\|\lambda_{i,t}-\lambda^*\|^2\nonumber\\
	&\leq\frac{N}{2}\|\lambda^*\|^2+\frac{1}{2}\sum\limits_{t=1}^T\alpha_t\left(\frac{1}{\gamma_t}-\frac{1}{\gamma_{t-1}}-\beta_t\right)\sum_{i=1}^N\|\lambda_{i,t}-\lambda^*\|^2.
\end{align*}
Then combining with the conditions in (\ref{equ55}), it can be verified that $\sum_{t=1}^{\infty}\mathcal{E}_t<\infty$.

By defining an auxiliary variable $\Theta_t:=\sum_{i=1}^ND_{\phi_i}(x_i^*,x_{i,t+1})+\sum_{s=t+1}^{\infty}\mathcal{E}_s$, it holds
\begin{align}
	\Theta_{t}\leq\Theta_{t-1}\leq\Theta_1.
\end{align}
By Doob's submartingale convergence theorem, it concludes that $\Theta_t$ converges to some finite value $\Theta_{\infty}$ and $\lim_{t\to\infty}\sum_{i=1}^ND_{\phi_i}(x_i^*,x_{i,t})=\Theta_{\infty}$.

Then, it suffices to prove the convergence of $x_t$ by the mathematical induction method. Assume that $x_t$ does not converge to $x^*$, then there is a subsequence of $\{x_t\}$, $\{x_{t_k}\}$, such that for some $\varepsilon>0$, $\|x_{t_k}-x^*\|\geq\varepsilon$. By (\ref{equ62}), one has
\begin{align}
	&\sum_{i=1}^ND_{\phi_i}(x_i^*,x_{i,t_{k+1}})\nonumber\\
	&\leq\sum_{i=1}^ND_{\phi_i}(x_i^*,x_{i,1})-\mu\sum_{s=1}^k\alpha_{t_s}\|x_{t_s}-x^*\|^2+\sum_{t=1}^{t_{k+1}-1}\mathcal{E}_t\nonumber\\
	&\leq\sum_{i=1}^ND_{\phi_i}(x_i^*,x_{i,1})-\mu\varepsilon\sum_{s=1}^k\alpha_{t_s}+\sum_{t=1}^{t_{k+1}-1}\mathcal{E}_t.\label{equ64}
\end{align}
Let $k\to\infty$, then $t_k\to\infty$. In view of $\sum_{t=1}^{\infty}\mathcal{E}_t<\infty$ and $\sum_{s=1}^{\infty}\alpha_{t_s}=\infty$, it can be derived from (\ref{equ64}) that
\begin{align}
	\lim_{k\to\infty}\sum_{i=1}^ND_{\phi_i}(x_i^*,x_{i,t_{k+1}})\leq-\infty,
\end{align}
which contradicts the finite limit of $\sum_{i=1}^ND_{\phi_i}(x_i^*,x_{i,t+1})$. Therefore, the hypothesis does not hold and hence $x_t$ converges to $x^*$.
\hfill$\blacksquare$

\subsection{Proof of Theorem \ref{thm3}}\label{E}

Summing over $t\in[T]$ on both sides of (\ref{equ62}), it can be derived that
\begin{align}
	&\mu\sum_{t=1}^T\|x_t-x^*\|^2\nonumber\\
	&\leq\sum_{t=1}^T\frac{1}{\alpha_t}\sum_{i=1}^N[D_{\phi_i}(x_{i}^*,x_{i,t})-D_{\phi_i}(x_{i}^*,x_{i,t+1})]\nonumber\\
	&~~~+\sum_{t=1}^T\frac{\mathcal{E}_t}{\alpha_t}.\label{equ75}
\end{align}
Replacing $\lambda$ and $\tilde{x}_i$ in (\ref{equ41}) and (\ref{equ43}) by $\lambda^*$ and $x_i^*$, respectively, one has from (\ref{equ75}) that
\begin{align}
	&\mu\sum_{t=1}^T\|x_t-x^*\|^2\nonumber\\
	&\leq\frac{2LK}{\alpha_{T+1}}+\frac{NM^2}{\mu_0}\sum_{t=1}^T\alpha_t+3\sum_{t=1}^Tc(t)+\frac{NM^2L^2}{\mu_0}\sum_{t=1}^T\frac{\alpha_t}{\beta_t^2}\nonumber\\
	&~~~+2NL^2\sum_{t=1}^T\gamma_t+\frac{N\Lambda}{2}\sum_{t=1}^T\beta_t+2L\sum_{t=1}^T\sum_{i=1}^NH_{i,t}\nonumber\\
	&~~~+\sum_{t=1}^T(\Lambda+\frac{L}{\beta_t})K_t+\frac{N\Lambda}{2}\nonumber\\
	&~~~+\frac{1}{2}\sum\limits_{t=1}^T\left(\frac{1}{\gamma_t}-\frac{1}{\gamma_{t-1}}-\beta_t\right)\sum_{i=1}^N\|\lambda_{i,t}-\lambda^*\|^2\nonumber\\
	&=\mathcal{O}(T^{\max\{b_1,1-b_1+2b_2,1-b_2,1-p,1-q+b_2\}}+1).\label{equ76}
\end{align}

By the convexity of $\|\cdot\|^2$, one has
\begin{align}
	\Big\|\frac{1}{T}\sum_{t=1}^Tx_t-x^*\Big\|^2\leq\frac{1}{T}\sum_{t=1}^T\|x_t-x^*\|^2.\label{equ77}
\end{align}
Therefore, combining (\ref{equ76}) and (\ref{equ77}) yields (\ref{equ3-}).
\hfill$\blacksquare$

\subsection{Proof of Theorem \ref{thm4}}\label{F}
The results in Lemma \ref{lemma1} still hold for Algorithm \ref{alg2} by only replacing $C_{i,t}$ in (\ref{equ21}) with $\hat{g}_{i,t}$.  Then, based on the optimality of $x_{i,t+1}$ in (\ref{equ82}), for any $x_i\in\Omega_i$, one has
\begin{align}
	&\left\langle{x}_{i,t+1}-x_i,\alpha_t\hat{\nabla}_iJ_{i,t}+\hat{\nabla}g_{i,t}\tilde{\lambda}_{i,t}\right\rangle\nonumber\\
	&+\left\langle{x}_{i,t+1}-x_i,\nabla\phi_i({x}_{i,t+1})-\nabla\phi_i({x}_{i,t})\right\rangle\leq 0,
\end{align}
which implies
\begin{align*}
	&\alpha_t\langle x_{i,t}-\tilde{x}_i,\nabla_iJ_{i,t}(x_t)\rangle\nonumber\\
	&\leq\alpha_t\langle\tilde{x}_i-x_{i,t},\hat{\nabla}_iJ_{i,t}-\nabla_iJ_{i,t}(x_t)\rangle\nonumber\\
	&~~~+\alpha_t\langle{x}_{i,t}-x_{i,t+1},\hat{\nabla}_iJ_{i,t}\rangle+\alpha_t\langle \tilde{x}_i-x_{i,t+1},\hat{\nabla}g_{i,t}\tilde{\lambda}_{i,t}\rangle\nonumber\\
	&~~~+D_{\phi_i}(\tilde{x}_i,x_{i,t})-D_{\phi_i}(\tilde{x}_i,{x}_{i,t+1})-\frac{\mu_0}{2}\|x_{i,t+1}-x_{i,t}\|^2
\end{align*}
by taking $x_i=\tilde{x}_i$,
where (\ref{equ10}) and (\ref{equ11}) have been used to derive this inequality.

Taking expectation on both sides of the above inequality over $\mathfrak{F}_t$ yields that
\begin{align}
	&\alpha_t\langle x_{i,t}-\tilde{x}_i,\nabla_iJ_{i,t}(x_t)\rangle\nonumber\\
	&\leq\alpha_t\langle\tilde{x}_i-x_{i,t},\mathbf{E}_{\mathfrak{F}_t}[\hat{\nabla}_iJ_{i,t}-\nabla_iJ_{i,t}(x_t)]\rangle\nonumber\\
	&~~~+\alpha_t\mathbf{E}_{\mathfrak{F}_t}[\langle{x}_{i,t}-x_{i,t+1},\hat{\nabla}_iJ_{i,t}\rangle]\nonumber\\
	&~~~+\alpha_t\mathbf{E}_{\mathfrak{F}_t}[\langle\tilde{x}_i-x_{i,t+1},\hat{\nabla}g_{i,t}\tilde{\lambda}_{i,t}\rangle]\nonumber\\
	&~~~+D_{\phi_i}(\tilde{x}_i,x_{i,t})-\mathbf{E}_{\mathfrak{F}_t}[D_{\phi_i}(\tilde{x}_i,{x}_{i,t+1})]\nonumber\\
	&~~~-\frac{\mu_0}{2}\mathbf{E}_{\mathfrak{F}_t}[\|x_{i,t+1}-x_{i,t}\|^2],\label{equ92}
\end{align}
where the fact that $x_{i,t}$ is independent of $\mathfrak{F}_t$ is applied.

For the first term on the right hand side of (\ref{equ92}), one has
\begin{align}
	&\alpha_t\langle\tilde{x}_i-x_{i,t},\mathbf{E}_{\mathfrak{F}_t}[\hat{\nabla}_iJ_{i,t}-\nabla_iJ_{i,t}(x_t)]\rangle\nonumber\\
	&\leq\alpha_t\|\tilde{x}_i-x_{i,t}\|\|\mathbf{E}_{\mathfrak{F}_t}[\hat{\nabla}_iJ_{i,t}-\nabla_iJ_{i,t}(x_t)]\|\nonumber\\
	&=\mathcal{O}(\alpha_t\bar{\delta}_{t}^2/\delta_{i,t}).
\end{align}
For the second term on the right hand side of (\ref{equ92}), it can be obtained that
\begin{align}
	&\alpha_t\mathbf{E}_{\mathfrak{F}_t}[\langle{x}_{i,t}-x_{i,t+1},\hat{\nabla}_iJ_{i,t}\rangle]\nonumber\\
	&\leq\frac{\mu_0}{4}\mathbf{E}[\|x_{i,t+1}-x_{i,t}\|^2]+\frac{\alpha_t^2}{\mu_0}\mathbf{E}[\|\hat{\nabla}_iJ_{i,t}\|^2].
\end{align}
For the third term on the right hand side of (\ref{equ92}), similar to (\ref{equ34}), one can derive that
\begin{align}
	&\alpha_t\mathbf{E}_{\mathfrak{F}_t}[\langle\tilde{x}_i-x_{i,t+1},\hat{\nabla}g_{i,t}\tilde{\lambda}_{i,t}\rangle]\nonumber\\
	&=\alpha_t\mathbf{E}_{\mathfrak{F}_t}[\langle \tilde{x}_i-x_{i,t+1},{\nabla}g_{i,t}(x_{i,t})\tilde{\lambda}_{i,t}\rangle]\nonumber\\
	&~~~+\alpha_t\mathbf{E}_{\mathfrak{F}_t}[\langle \tilde{x}_i-x_{i,t+1},(\hat{\nabla}g_{i,t}-{\nabla}g_{i,t}(x_{i,t}))\tilde{\lambda}_{i,t}\rangle]\nonumber\\
	&\leq4\sqrt{N}L^2\alpha_t\sum\limits_{s=0}^{t-1}\sigma^{s}\gamma_{t-1-s}+\alpha_t\overline{\lambda}_{t}^{\top}(g_{i,t}(\tilde{x}_{i})-g_{i,t}(x_{i,t}))\nonumber\\
	&~~~+\frac{L^2M^2}{\mu_0}\frac{\alpha_t^2}{\beta_t^2}+\frac{\mu_0}{4}\mathbf{E}_{\mathfrak{F}_t}[\|x_{i,t}-{x}_{i,t+1}\|^2]\nonumber\\
	&~~~+2L^2\alpha_t\mathbf{E}_{\mathfrak{F}_t}[\|\hat{\nabla}g_{i,t}-{\nabla}g_{i,t}(x_{i,t})\|]/\beta_t.\label{equ95}
\end{align}

By Lemma \ref{lemma4}, combining with the proofs of Lemma \ref{lemma3} and Theorem \ref{thm1}, one has from (\ref{equ92})--(\ref{equ95}) that
\begin{align}
	&\sum_{t=1}^T\mathbf{E}[\langle x_{i,t}-\tilde{x}_i,\nabla_iJ_{i,t}(x_t)\rangle]\nonumber\\
	&=\mathcal{O}\Big(\sum_{t=1}^T(\delta_{i,t}+\gamma_t+\frac{\alpha_t}{\delta_{i,t}}+\frac{\alpha_t}{\beta_t^2}+\frac{\delta_t}{\beta_t})+\frac{1}{\alpha_T}\Big)\nonumber\\
	&~~~+\sum_{t=1}^T\mathbf{E}[\overline{\lambda}_t^{\top}(\hat{g}_{t}-g_t(x_t))]-\lambda^{\top}\mathbf{E}[\sum_{t=1}^T\hat{g}_{t}]\nonumber\\
	&~~~+\frac{N}{2}(1+\sum_{t=1}^T\beta_t)\|\lambda\|^2\nonumber\\
	&~~~+\frac{1}{2}\sum_{t=1}^T\Big(\frac{1}{\gamma_t}-\frac{1}{\gamma_{t-1}}-\beta_t\Big)\sum_{i=1}^N\mathbf{E}[\|\lambda_{i,t}-\lambda\|^2].\label{equ96}
\end{align}
Note that
\begin{align}
	\sum_{t=1}^T\overline{\lambda}_t^{\top}\Big(\hat{g}_{t}-g_t(x_t)\Big)
	&\leq\sum_{t=1}^T\|\overline{\lambda}_t\|\Big\|\hat{g}_{t}-g_t(x_t)\Big\|\nonumber\\
	&=\mathcal{O}(\sum_{t=1}^T\delta_t/\beta_t).\label{}
\end{align}
By setting $\lambda={\bf 0}_m$, one has
\begin{align}
	&\sum_{t=1}^T\mathbf{E}[\langle x_{i,t}-\tilde{x}_i,\nabla_iJ_{i,t}(x_t)\rangle]\nonumber\\
	&=\mathcal{O}\Big(\sum_{t=1}^T(\delta_{i,t}+\gamma_t+\frac{\alpha_t}{\delta_{i,t}}+\frac{\alpha_t}{\beta_t^2}+\frac{\delta_t}{\beta_t})+\frac{1}{\alpha_T}\Big).
\end{align}
Since
\begin{align}
	&\mathbf{E}[Reg_i(T)]\nonumber\\
	&\leq\mathbf{E}\Big[\sum_{t=1}^T\langle \hat{x}_{i,t}-\tilde{x}_i,\nabla_iJ_{i,t}(\hat{x}_{t})\rangle\Big]\nonumber\\
	&=\mathbf{E}\Big[\sum_{t=1}^T\langle x_{i,t}-\tilde{x}_i,\nabla_iJ_{i,t}({x}_{t})\rangle\Big]\nonumber\\
	&~~~+\mathbf{E}\Big[\sum_{t=1}^T\langle \hat{x}_{i,t}-x_{i,t},\nabla_iJ_{i,t}({x}_{t})\rangle\Big]\nonumber\\
	&~~~+\mathbf{E}\Big[\sum_{t=1}^T\langle \hat{x}_{i,t}-\tilde{x}_i,\nabla_iJ_{i,t}(\hat{x}_{t})-\nabla_iJ_{i,t}({x}_{t})\rangle\Big]\nonumber\\
	&=\mathbf{E}\Big[\sum_{t=1}^T\langle x_{i,t}-\tilde{x}_i,\nabla_iJ_{i,t}({x}_{t})\rangle\Big]+\mathcal{O}(\sum_{t=1}^T\delta_{i,t}),
\end{align}
 it can be concluded that (\ref{equ89}) holds based on the selection of the stepsizes.

Let $\lambda=\frac{2\mathbf{E}\left[\left[\sum_{t=1}^T\hat{g}_t\right]_+\right]}{B_3(T)}$,  where $B_3(T)$ is defined in Lemma \ref{lemma3}.  Then, it can be derived from (\ref{equ96}) that
\begin{align}
	&\sum_{t=1}^T\mathbf{E}[\langle x_{i,t}-\tilde{x}_i,\nabla_iJ_{i,t}(x_t)\rangle]\nonumber\\
	&=\mathcal{O}\Big(\sum_{t=1}^T(\delta_{i,t}+\gamma_t+\frac{\alpha_t}{\delta_{i,t}}+\frac{\alpha_t}{\beta_t^2}+\frac{\delta_t}{\beta_t})+\frac{1}{\alpha_T}\Big)\nonumber\\
	&~~~-\frac{(\mathbf{E}[R_g(T)])^2}{B_3(T)}.
\end{align}

By (\ref{equ53}) and $\langle x_{i,t}-\tilde{x}_i,\nabla_iJ_{i,t}(x_t)\rangle\geq0$, it can be obtained that
\begin{align*}
	&(\mathbf{E}[R_g(T)])^2\\	&=\mathcal{O}\Big(B_3(T)\sum_{t=1}^T(\delta_{i,t}+\gamma_t+\frac{\alpha_t}{\delta_{i,t}}+\frac{\alpha_t}{\beta_t^2}+\frac{\delta_t}{\beta_t})+\frac{B_3(T)}{\alpha_T}\Big).
\end{align*}
Via a simple computation, the bound on $\mathbf{E}[R_g(T)]$ can be derived as in (\ref{equ90}).
\hfill$\blacksquare$

\end{appendix}
\bibliographystyle{IEEEtran}
\bibliography{MinMeng}

\begin{thebibliography}{10}
\providecommand{\url}[1]{#1}
\csname url@samestyle\endcsname
\providecommand{\newblock}{\relax}
\providecommand{\bibinfo}[2]{#2}
\providecommand{\BIBentrySTDinterwordspacing}{\spaceskip=0pt\relax}
\providecommand{\BIBentryALTinterwordstretchfactor}{4}
\providecommand{\BIBentryALTinterwordspacing}{\spaceskip=\fontdimen2\font plus
\BIBentryALTinterwordstretchfactor\fontdimen3\font minus
  \fontdimen4\font\relax}
\providecommand{\BIBforeignlanguage}[2]{{%
\expandafter\ifx\csname l@#1\endcsname\relax
\typeout{** WARNING: IEEEtran.bst: No hyphenation pattern has been}%
\typeout{** loaded for the language `#1'. Using the pattern for}%
\typeout{** the default language instead.}%
\else
\language=\csname l@#1\endcsname
\fi
#2}}
\providecommand{\BIBdecl}{\relax}
\BIBdecl

\bibitem{ghaderi2014opinion}
J.~Ghaderi and R.~Srikant, ``{Opinion dynamics in social networks with stubborn
  agents: Equilibrium and convergence rate},'' \emph{Automatica}, vol.~50,
  no.~12, pp. 3209--3215, 2014.

\bibitem{stankovic2012distributed}
M.~S. Stankovic, K.~H. Johansson, and D.~M. Stipanovic, ``{Distributed seeking
  of Nash equilibria with applications to mobile sensor networks},'' \emph{IEEE
  Transactions on Automatic Control}, vol.~57, no.~4, pp. 904--919, 2012.

\bibitem{saad2012game}
W.~Saad, Z.~Han, H.~V. Poor, and T.~Ba\c{s}ar, ``Game-theoretic methods for the
  smart grid: An overview of microgrid systems, demand-side management, and
  smart grid communications,'' \emph{IEEE Signal Processing Magazine}, vol.~29,
  no.~5, pp. 86--105, 2012.

\bibitem{basar1999dynamic}
T.~Ba\c{s}ar and G.~J. Olsder, \emph{{Dynamic Noncooperative Game
  Theory}}.\hskip 1em plus 0.5em minus 0.4em\relax SIAM, 1999, vol.~23.

\bibitem{facchinei2010generalized}
F.~Facchinei and C.~Kanzow, ``{Generalized Nash equilibrium problems},''
  \emph{Annals of Operations Research}, vol. 175, no.~1, pp. 177--211, 2010.

\bibitem{yu2017distributed}
C.~K. Yu, M.~Van Der~Schaar, and A.~H. Sayed, ``{Distributed learning for
  stochastic generalized Nash equilibrium problems},'' \emph{IEEE Transactions
  on Signal Processing}, vol.~65, no.~15, pp. 3893--3908, 2017.

\bibitem{shamma2005dynamic}
J.~S. Shamma and G.~Arslan, ``{Dynamic fictitious play, dynamic gradient play,
  and distributed convergence to Nash equilibria},'' \emph{IEEE Transactions on
  Automatic Control}, vol.~50, no.~3, pp. 312--327, 2005.

\bibitem{de2019distributed}
C.~De~Persis and S.~Grammatico, ``{Distributed averaging integral Nash
  equilibrium seeking on networks},'' \emph{Automatica}, vol. 110, p. 108548,
  2019.

\bibitem{gadjov2019passivity}
D.~Gadjov and L.~Pavel, ``{A passivity-based approach to Nash equilibrium
  seeking over networks},'' \emph{IEEE Transactions on Automatic Control},
  vol.~64, no.~3, pp. 1077--1092, 2019.

\bibitem{koshal2016distributed}
J.~Koshal, A.~Nedi{\'c}, and U.~V. Shanbhag, ``Distributed algorithms for
  aggregative games on graphs,'' \emph{Operations Research}, vol.~64, no.~3,
  pp. 680--704, 2016.

\bibitem{salehisadaghiani2019distributed}
F.~Salehisadaghiani, W.~Shi, and L.~Pavel, ``{Distributed Nash equilibrium
  seeking under partial-decision information via the alternating direction
  method of multipliers},'' \emph{Automatica}, vol. 103, pp. 27--35, 2019.

\bibitem{tatarenko2019geometric}
T.~Tatarenko and A.~Nedi{\'c}, ``Geometric convergence of distributed gradient
  play in games with unconstrained action sets,'' \emph{IFAC-PapersOnLine},
  vol.~53, no.~2, pp. 3367--3372, 2020.

\bibitem{cesa2006prediction}
N.~Cesa-Bianchi and G.~Lugosi, \emph{{Prediction, Learning, and Games}}.\hskip
  1em plus 0.5em minus 0.4em\relax Cambridge University Press, 2006.

\bibitem{abernethyoptimal}
J.~Abernethy, P.~L. Bartlett, A.~Rakhlin, and A.~Tewari, ``Optimal strategies
  and minimax lower bounds for online convex games,'' in \emph{COLT'08:
  Proceedings of the 21st Annual Conference on Learning Theory}, 2008.

\bibitem{duvocelle2022multiagent}
B.~Duvocelle, P.~Mertikopoulos, M.~Staudigl, and D.~Vermeulen, ``Multiagent
  online learning in time-varying games,'' \emph{Mathematics of Operations
  Research}, 2022.

\bibitem{lu2020online}
K.~Lu, H.~Li, and L.~Wang, ``{Online distributed algorithms for seeking
  generalized Nash equilibria in dynamic environments},'' \emph{IEEE
  Transactions on Automatic Control}, vol.~66, no.~5, pp. 2289--2296, 2021.

\bibitem{viossat2013no}
Y.~Viossat and A.~Zapechelnyuk, ``No-regret dynamics and fictitious play,''
  \emph{Journal of Economic Theory}, vol. 148, no.~2, pp. 825--842, 2013.

\bibitem{mertikopoulos2018cycles}
P.~Mertikopoulos, C.~Papadimitriou, and G.~Piliouras, ``Cycles in adversarial
  regularized learning,'' in \emph{Proceedings of the Twenty-Ninth Annual
  ACM-SIAM Symposium on Discrete Algorithms}.\hskip 1em plus 0.5em minus
  0.4em\relax SIAM, 2018, pp. 2703--2717.

\bibitem{heliou2017learning}
A.~Heliou, J.~Cohen, and P.~Mertikopoulos, ``Learning with bandit feedback in
  potential games,'' \emph{Advances in Neural Information Processing Systems},
  vol.~30, 2017.

\bibitem{cao2019online}
X.~Cao and K.~J.~R. Liu, ``Online convex optimization with time-varying
  constraints and bandit feedback,'' \emph{IEEE Transactions on Automatic
  Control}, vol.~64, no.~7, pp. 2665--2680, 2019.

\bibitem{yi2020distributed_b}
X.~Yi, X.~Li, T.~Yang, L.~Xie, T.~Chai, and K.~H. Johansson, ``Distributed
  bandit online convex optimization with time-varying coupled inequality
  constraints,'' \emph{IEEE Transactions on Automatic Control}, 2020, DOI:
  10.1109/TAC.2020.3030883.

\bibitem{bravo2018bandit}
M.~Bravo, D.~Leslie, and P.~Mertikopoulos, ``Bandit learning in concave
  $n$-person games,'' in \emph{Proceedings of the 32nd International Conference
  on Neural Information Processing Systems (NeurIPS)}, 2018, pp. 5666--5676.

\bibitem{mertikopoulos2019learning}
P.~Mertikopoulos and Z.~Zhou, ``Learning in games with continuous action sets
  and unknown payoff functions,'' \emph{Mathematical Programming}, vol. 173,
  no.~1, pp. 465--507, 2019.

\bibitem{li2020}
X.~Li, L.~Xie, and Y.~Hong, ``Distributed aggregative optimization over
  multi-agent networks,'' \emph{IEEE Transactions on Automatic Control},
  vol.~67, no.~6, pp. 3165--3171, 2022.

\bibitem{salehisadaghiani2016distributed}
F.~Salehisadaghiani and L.~Pavel, ``{Distributed Nash equilibrium seeking: A
  gossip-based algorithm},'' \emph{Automatica}, vol.~72, pp. 209--216, 2016.

\bibitem{bregman1967relaxation}
L.~M. Bregman, ``The relaxation method of finding the common point of convex
  sets and its application to the solution of problems in convex programming,''
  \emph{USSR Computational Mathematics and Mathematical Physics}, vol.~7,
  no.~3, pp. 200--217, 1967.

\bibitem{hall2015online}
E.~C. Hall and R.~M. Willett, ``Online convex optimization in dynamic
  environments,'' \emph{IEEE Journal of Selected Topics in Signal Processing},
  vol.~9, no.~4, pp. 647--662, 2015.

\bibitem{neely2017online}
M.~J. Neely and H.~Yu, ``Online convex optimization with time-varying
  constraints,'' \emph{arXiv preprint arXiv:1702.04783}, 2017.

\bibitem{facchinei2009nash}
F.~Facchinei and J.-S. Pang, ``Nash equilibria: The variational approach,''
  \emph{Convex Optimization in Signal Processing and Communications}, pp.
  443--449, 2010.

\bibitem{liang2017distributed}
S.~Liang, P.~Yi, and Y.~Hong, ``{Distributed Nash equilibrium seeking for
  aggregative games with coupled constraints},'' \emph{Automatica}, vol.~85,
  pp. 179--185, 2017.

\bibitem{pavel2020distributed}
L.~Pavel, ``{Distributed GNE seeking under partial-decision information over
  networks via a doubly-augmented operator splitting approach},'' \emph{IEEE
  Transactions on Automatic Control}, vol.~65, no.~4, pp. 1584--1597, 2020.

\bibitem{kulkarni2012variational}
A.~A. Kulkarni and U.~V. Shanbhag, ``{On the variational equilibrium as a
  refinement of the generalized Nash equilibrium},'' \emph{Automatica},
  vol.~48, no.~1, pp. 45--55, 2012.

\bibitem{nedic2009approximate}
A.~Nedi{\'c} and A.~Ozdaglar, ``Approximate primal solutions and rate analysis
  for dual subgradient methods,'' \emph{SIAM Journal on Optimization}, vol.~19,
  no.~4, pp. 1757--1780, 2009.

\bibitem{yi2020distributed}
X.~Yi, X.~Li, L.~Xie, and K.~H. Johansson, ``Distributed online convex
  optimization with time-varying coupled inequality constraints,'' \emph{IEEE
  Transactions on Signal Processing}, vol.~68, pp. 731--746, 2020.

\end{thebibliography}

\end{document}